%% file: IEEE-main.tex
\documentclass[conference]{IEEEtran}
\renewcommand{\IEEEQED}{\IEEEQEDopen}

\usepackage{cite}
\usepackage{amsmath,amssymb,amsfonts}
\usepackage{algorithm}
\usepackage{algpseudocode}
\usepackage{textcomp}
\usepackage{xcolor}
\usepackage{graphicx}
\usepackage{booktabs}
\usepackage{tabularx}
\usepackage[switch,mathlines]{lineno}
\usepackage{makecell}
\usepackage[hidelinks]{hyperref}

\newtheorem{definition}{Definition}

\newtheorem{lemma}{Lemma}
\newtheorem{theorem}{Theorem}

\DeclareMathOperator{\Op}{op}

\newcommand{\changed}[1]{{#1}}

\def\BibTeX{{\rm B\kern-.05em{\sc i\kern-.025em b}\kern-.08em
    T\kern-.1667em\lower.7ex\hbox{E}\kern-.125emX}}
\begin{document}

\title{Compilation of Generalized Matrix Chains with Symbolic Sizes
}

\author{\IEEEauthorblockN{Francisco López}
\IEEEauthorblockA{Department of Computing Science \\
Umeå Universitet \\
Umeå, Sweden \\
flopz@cs.umu.se}
\and
\IEEEauthorblockN{Lars Karlsson}
\IEEEauthorblockA{Department of Computing Science \\ 
Umeå Universitet \\
Umeå, Sweden \\
larsk@cs.umu.se}
\and
\IEEEauthorblockN{Paolo Bientinesi}
\IEEEauthorblockA{Department of Computing Science \\
Umeå Universitet \\
Umeå, Sweden \\
pauldj@cs.umu.se}
}

\maketitle
\thispagestyle{plain}
\pagestyle{plain}

% CRITICAL: Do Not Use Symbols, Special Characters, Footnotes, or Math in Paper Title or Abstract.
\begin{abstract}
    Generalized Matrix Chains (GMCs) are products of matrices where each matrix carries features (e.g., general, symmetric, triangular, positive-definite) and is optionally transposed and/or inverted.
    GMCs are commonly evaluated via sequences of calls to BLAS and LAPACK kernels.
    When matrix sizes are known, one can craft a sequence of kernel calls to evaluate a GMC that minimizes some cost, e.g., the number of floating-point operations (FLOPs).
    Even in these circumstances, high-level languages and libraries, upon which users usually rely, typically perform a suboptimal mapping of the input GMC onto a sequence of kernels.
    In this work, we go one step beyond and consider matrix sizes to be symbolic (unknown); this changes the nature of the problem since no single sequence of kernel calls is optimal for all possible combinations of matrix sizes.
    We design and evaluate a code generator for GMCs with symbolic sizes that relies on multi-versioning.
    At compile-time, when the GMC is known but the sizes are not, code is generated for a few carefully selected sequences of kernel calls.
    At run-time, when sizes become known, the best generated variant for the matrix sizes at hand is selected and executed.
    The code generator uses new theoretical results that guarantee that the cost is within a constant factor from optimal for all matrix sizes and an empirical tuning component that further tightens the gap to optimality in practice.
    In experiments, we found that the increase above optimal in both FLOPs and execution time of the generated code was less than 15\% for 95\% of the tested chains.
\end{abstract}


\input{sections/1_intro}
\input{sections/2_gmc}

\input{sections/4_algorithm_generation}

\input{sections/5_theory}

\input{sections/6_expansion}
\input{sections/7_experiments}

\input{sections/8_conclusions}

\section*{Acknowledgment}
This research was conducted using the resources of High Performance Computing Center North (HPC2N).
We thank Conrad Sanderson and Ryan Curtin for helpful discussions on how matrix chains are evaluated in Armadillo.

\bibliographystyle{IEEEtran}
\bibliography{references}

\newpage
\appendix
\input{sections/appendix_proof}
\input{sections/appendix_kernels}

\end{document}

%% file: sections/1_intro.tex
\section{Introduction}

Despite the significant effort put into the development of high-performance matrix kernels by the numerical linear algebra community, users rarely undertake the time-consuming and error-prone process of directly invoking such kernels.
In fact, high-level languages and libraries are ever more popular, although they typically perform poor mappings of linear algebra expressions to sequences of kernel calls, resulting in subpar performance~\cite{psarras2022linear}.
The situation is even more challenging when the sizes of the matrices are unknown at compile time; in this case, one cannot rely on one single mapping, since no mapping performs well on the entire space of matrix sizes. 

A well-known example is the classic \emph{Matrix Chain Problem} (MCP), where the optimal parenthesization (i.e., sequence of calls to a multiplication kernel) depends on the matrix sizes.
For example, for column vectors with $m$ elements, the parenthesization $x^T (y z^T)$ performs $m$ times more multiplications than $(x^T y)z^T$.
%Finding an efficient mapping to kernels becomes even more convoluted when the sizes of the operands in the expressions are unknown.
In this paper, we present and evaluate a code generator for a large class of expressions known as generalized matrix chains.
In contrast to existing solutions, we consider the important case of unknown matrix sizes at compile-time.
\changed{This problem is commonly encountered in practice. As examples, the Kalman filter~\cite{rao2017robust} and the Tikhonov regularization~\cite{noschese2016some} are widely used tools that are computed through a linear algebra expression. They appear in a multitude of engineering and data-science applications, and while their expression is fixed, the size of the operands varies in different contexts and oftentimes becomes known only at run-time. }

The problem of translating an expression into a sequence of kernel calls is known as the \emph{Linear Algebra Mapping Problem} (LAMP), formally defined as follows~\cite{psarras2022linear}.
\begin{definition}[Linear Algebra Mapping Problem]
    Given a linear algebra expression $\mathcal L$, a set of instructions $\mathcal I$, and a cost function $\mathcal C$, construct a program $\mathcal P$ using the instructions in $\mathcal I$ that computes $\mathcal L$ while minimizing $\mathcal C(\mathcal P)$.
\end{definition}

The MCP, which has been extensively studied (see Section~\ref{sec:related_work}), is captured by the instance of the LAMP where the expressions in $\mathcal L$ are matrix products $M_1 M_2 \cdots M_n$ with $M_i$ being a matrix of size $q_{i-1} \times q_i$ (i.e., standard matrix chains), $\mathcal I$ only contains an instruction that computes the matrix product (e.g., \textsc{gemm} in BLAS~\cite{dongarra1990set}), and $\mathcal C$ is the number of floating-point operations (FLOPs).
A solution to the MCP is a \emph{parenthesization} that minimizes the number of FLOPs~\cite{godbole1973efficient}.

Standard matrix chains are rare in practice~\cite{barthels2018generalized}.
By contrast, \emph{Generalized Matrix Chains} (GMCs), where matrices have features (e.g., symmetry) and can be transposed and/or inverted, are much more common.
For example, the GMC $G_{1} L_{1}^{-1} G_{2} L_{2}^{-1}$, where $L_{1}$ and $L_{2}$ are triangular, appears in a blocked algorithm for the inversion of a triangular matrix~\cite{bientinesi2006mechanical},
while $G_{1} G_2 G_3^T M^{-1}$ appears in the ensemble Kalman filter~\cite{rao2017robust}.
Many more examples can be found in the context of computer vision~\cite{bronstein2016consistent}, optimization~\cite{straszak2015natural}, information theory~\cite{albataineh2014blind,hejazi2015robust}, signal processing~\cite{ding2016sparsity,nino2019parallel}, regularization~\cite{noschese2016some}, and the simulation of power grids~\cite{ronellenfitsch2016dual}.

The \emph{Generalized Matrix Chain Problem} (GMCP)~\cite{barthels2018generalized}, is the instance of the LAMP in which $\mathcal L$, $\mathcal I$, and $\mathcal C$ are chosen as follows.
The expressions in $\mathcal L$ are of the form
\begin{equation}\label{eq:gmc}
  \Op(M_1) \Op(M_2) \cdots \Op(M_n),
\end{equation}
where $M_i$ has size $q_{i-1} \times q_i$, may exhibit different features (e.g., symmetric, triangular, positive-definite), and can be subjected to unary operators: $\Op(M) = M, M^T, M^{-1}, M^{-T}$.
The instruction set $\mathcal I$ contains kernels for multiplying two matrices, solving linear systems, and inverting matrices, such as those provided by the BLAS~\cite{lawson1979basic,dongarra1985proposal,dongarra1990set} and LAPACK~\cite{anderson1999lapack} libraries.
The cost function $\mathcal C$ can be, for example, the number of FLOPs or the execution time.
A solution to the GMCP is a sequence of kernel calls that minimizes $\mathcal{C}$.

% The problem we consider: GMCP with symbolic sizes
In most applications, matrix sizes are rarely known at compile-time; symbolic sizes are the norm.
Here we consider the GMCP with symbolic sizes.
At compile-time, the shape (i.e., features and unary operators acting upon matrices) of the chain is given but the sizes are symbolic.
The goal is to generate code for the given shape that can efficiently evaluate any instance of the symbolic chain.
We tackle two related versions of the problem, namely, (a) when $\mathcal C$ is the number of FLOPs, and (b) when $\mathcal C$ is the execution time.

% How the landscape changes when we move from fixed to symbolic sizes.
Moving from fixed---i.e., known at compile time---to symbolic sizes radically changes the nature of the code generation problem.
With fixed sizes, there is an optimal sequence of kernel calls.
Compiling for fixed sizes therefore boils down to finding and generating code for that best sequence.
However, with symbolic sizes, different sequences can be best in different regions of the instance space~\cite{lopez2025parenth}.
In fact, a sequence that is best in one region can be arbitrarily far from optimal in another~\cite{lopez2025parenth}.
\changed{
Hence, generating code for just one sequence gives no performance guarantees.
A natural alternative is to generate code for all possible sequences at compile-time and then dispatch to the best one for a given instance at run-time. 
However, the number of sequences grows exponentially with the length of the chain, which makes the overheads both in terms of code size and run-time for dispatch prohibitively expensive.
}
% Both the code size overhead and the run-time overhead for variant dispatch makes this approach expensive. % to generate code for all possible code variants.

%A compiler for symbolic sizes must therefore generate a carefully chosen set of sequences to guarantee decent performance everywhere. 

% This says: we do nothing at compile-time, and we do everything at run-time.
% Here we want to motivate that doing everything at run-time is bad.

\changed{
A very different alternative is to search for an optimal sequence at run-time when the sizes become known and then immediately execute it.
This does not involve any code generation at all.
For a regular matrix chain, the search can be done with the classical dynamic programming algorithm and the execution boils down to repeatedly calling \textsc{gemm}.
For generalized matrix chains, this approach becomes significantly more complicated.
The search for an optimal sequence requires inferring features (e.g., lower-triangular) of intermediate results, assigning an appropriate kernel to each operation depending on the operand features, and potentially rewriting expressions involving transpositions and/or inversions.
The execution of the optimal sequence needs to dispatch to more than a dozen different kernels, each with different configurations (e.g., operand transposition patterns) and manage memory accordingly.
In summary, both the search and the sequence execution are non-trivial tasks for GMCs and the corresponding software adds a non-negligible code-size overhead.
Moreover, to the best of our knowledge, the only complete solution to the GMCP is Linnea~\cite{barthels2021linnea}.
However, Linnea was not designed for efficiency and cannot be used in low-latency applications such as online expression evaluation.

% Another alternative is to use an interpretive approach. 
% No variants are generated at compile-time.
% Instead, a \emph{planner} could search for an optimal variant at runtime once the sizes are known.
% The optimal plan would then be executed by an \emph{interpreter} that calls the kernels and manages memory according to the plan.
% For a regular matrix chain, the planner simply needs to find an optimal parenthesization (e.g., using the classical dynamic programming algorithm), and the interpreter repeatedly calls the \textsc{gemm} kernel. 
% For a generalized matrix chain, both the planner and the interpreter become significantly more complicated. 
% The planner needs to deduce features of intermediate results, assign a kernel to each operation, and potentially rewrite expressions involving transpositions and/or inversions. 
% The interpreter needs to dispatch to more than a dozen different kernels, each with different configurations (e.g., operands transposition patterns).
% Both the planner and the interpreter add significant code size overhead as well as run-time overhead.
% Most of the overhead is caused by the planner.
% Hence, an interpretive approach has considerable drawbacks.
}

% what we present here: code generator for gmc with symbolic sizes
We propose and evaluate an approach to the symbolic compilation problem based on \emph{multi-versioning}~\cite{cooper1992procedure}.
At compile-time, a few code variants are generated along with a dispatch function that at run-time selects the best variant for a given instance of the symbolic chain.
We propose a rigorous approach based on new theoretical results that provide performance guarantees.
Combined with a tuning mechanism that allows the user to control the trade-off between overhead and performance, we end up with a practical code generator for GMCs with symbolic sizes.

\begin{figure}[htbp]
  \centering
  \includegraphics[width=\columnwidth]{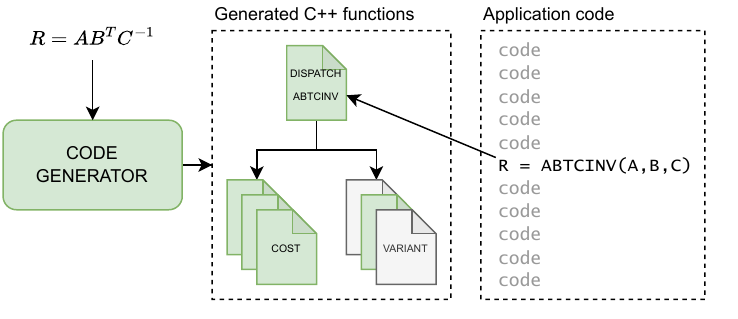}
  \caption{A multi-versioning code generator for compiling generalized matrix chains with symbolic sizes.}\label{fig:setup}
\end{figure}

\figurename~\ref{fig:setup} illustrates the setup. 
The shape of the chain, specified using the grammar in \figurename~\ref{fig:grammar_input}, is the input to the code generator, which then produces the following:
\begin{itemize}
  \item Code for a set of C\texttt{++} functions that implement a few different variants paired with C\texttt{++} functions that estimate the associated cost (FLOPs or execution time) given concrete matrix sizes.
  \item Code for a C\texttt{++} function that dispatches control to the best variant for a given combination of matrix sizes.
\end{itemize}
The generated code is compiled and linked to the user's application code.
At run-time, the application calls the dispatch function with concrete matrices as arguments. %and supplies it with matrix sizes as well as matrices to operate upon.
The dispatch function evaluates the cost of every variant before passing control to the best one for the given matrix sizes.
An application can contain multiple sets of generated code: One for each type of generalized matrix chain used by the application.

With this article, we make the following contributions:
\begin{itemize}
  \item \changed{We propose and evaluate the first approach to compile generalized matrix chains with symbolic sizes based on a theoretically-grounded application of multi-versioning.}
  \item \changed{We present novel theoretical results that show how to select at most $n+1$ ($n$ is the number of matrices) code variants for any given GMC such that their cost is within a constant factor from optimal for all combinations of matrix sizes.}
  \item We present and evaluate a procedure to incrementally expand sets of code variants to improve performance at the expense of increased code size overhead.
\end{itemize}
A code generator for generalized matrix chains with symbolic sizes is one step towards a linear algebra compiler for more general symbolic expressions, which so far remains an unsolved problem.

\emph{Organization of the paper}.
Section~\ref{sec:related_work} presents the related work.
Section~\ref{sec:gmc} presents the design of our code generator for the GMCP.\@
Section~\ref{sec:gen_variants} describes how the code generator constructs a code variant from a parenthesization.
Section~\ref{sec:theory} presents novel theoretical results that guide the selection of a small set of parenthesizations for which to generate variants.
Section~\ref{sec:expansion} presents a procedure to expand sets of variants.
Section~\ref{sec:experiments} presents numerical experiments that evaluate the effectiveness of the code generator.
Section~\ref{sec:conclusion} concludes and outlines future work.

\begin{figure}[htbp]
  \centering{}
  \begin{align*}
    \text{program} &\rightarrow \text{definitions}\ \ \text{expression}\\
    \text{definitions} &\rightarrow \text{definition}^+  \\
    \text{definition} &\rightarrow \textbf{Matrix}\ \text{ident}\ \left< \text{structure},\text{property} \right>; \\
    \text{structure} &\rightarrow \textbf{General}\ |\ \textbf{Symmetric}\ |\ \textbf{LowerTri}\ |\ \ldots \\
    \text{property} &\rightarrow  \textbf{Singular}\ |\ \textbf{SPD}\ |\ \textbf{Orthogonal}\ |\ \ldots \\
    \text{expression} &\rightarrow \text{lhs} := \text{operand}\ {\{*\ \text{operand} \}}^{+}; \\
    \text{operand} &\rightarrow \text{ident}\ |\ \text{ident}^{T}\ |\ \text{ident}^{-1}\ |\ \text{ident}^{-T} \\
    \text{ident} & \rightarrow \textbf{A} \ |\ \textbf{B}\ |\ \ldots
  \end{align*}
  \caption{Grammar for the code generator's input.}\label{fig:grammar_input}
\end{figure}

\section{Related Work}\label{sec:related_work}
The MCP has been extensively studied ever since Godbole first formulated it~\cite{godbole1973efficient}.
Numerous exact~\cite{godbole1973efficient,hu1982computation,hu1984computation} and approximate~\cite{chandra1975computing,chin1978n,hu1981n,lopez2025parenth} algorithms have been published throughout the years.
Some works provide parallel algorithms for the solution of the MCP~\cite{bradford1998efficient,ramanan1996efficient,strate1990parallelization}, while others solve the MCP sequentially, but target parallel systems~\cite{lee2003processor} or accelerators/GPUs~\cite{nishida2011accelerating}.

The GMCP was introduced by Barthels et al.~\cite{barthels2018generalized}.
They also presented a dynamic programming algorithm to solve GMCs with known matrix sizes.
That same algorithm was later used in Linnea~\cite{barthels2021linnea}, a compiler for the automatic generation of optimized code for more general linear algebra expressions with concrete sizes through invocations to BLAS
and LAPACK~\cite{anderson1999lapack} (and some bespoke) kernels.
Given that BLAS and LAPACK kernels rarely offer optimal performance for every combination of operation and sizes, other linear algebra compilers generate code that relies on loop nests instead of standard kernels.
Examples include Build to Order~\cite{siek2008build}, which focuses on bandwidth-bound operations (corresponding to BLAS 1 and 2), and LG\textsc{en}~\cite{spampinato2016basic} and SL\textsc{in}G\textsc{en}~\cite{spampinato2018program}, which focus on operations upon small matrices.

GMCs are commonly evaluated via high-level languages and environments such as Matlab, Octave, R, Julia, NumPy, PyTorch, and TensorFlow~\cite{psarras2022linear,9835658}. 
These languages allow expressions to be input in a form that closely resembles mathematical notation.
The language's compiler or interpreter then automatically maps the expression to kernels.
Unfortunately, these automatic mappings rarely yield efficient evaluations~\cite{psarras2022linear,9835658}.
In MATLAB, for example, products are evaluated left-to-right~\cite{mathworks2025doc} regardless of the matrix sizes.
Moreover, when users input \texttt{X=inv(A)*B}, the matrix $A$ will be explicitly inverted even though solving the linear system $AX = B$ is mathematically equivalent, faster, and numerically more stable.
By contrast, in Julia, where types are used to represent a small set of basic properties in conjunction with multiple dispatch, standard matrix chains are optimally parenthesized.

MATLAB offers an add-on, called Coder, that can generate C/C\texttt{++} code from M-files.
Linear algebra expressions are mapped to BLAS and LAPACK kernels and sizes can be symbolic at compile-time.
Thus, Coder technically offers a solution to the GMCP with symbolic sizes.
However, the generated code always evaluates the chain left-to-right, which can be arbitrarily far from optimal~\cite{chandra1975computing,lopez2025parenth}.

There also exist templated C\texttt{++} libraries that allow for high-level input of linear algebra expressions, such as Blaze~\cite{iglberger2012expression}, Blitz++~\cite{veldhuizen1998arrays}, Eigen~\cite{jacob2010eigen}, and Armadillo~\cite{sanderson2025armadillo}.
Expressions are mostly evaluated following simple rules, such as left-to-right evaluation, that usually yield suboptimal performance.
Armadillo is, however, the exception in this group.
It includes a heuristics-based approximate solution to the MCP and a more advanced mapping of fixed-sized expressions to kernels. 
Users of these libraries must commonly specify how the inverse operator should be translated into operations, as in MATLAB, and matrix features are represented through types, as in Julia.

In summary, there is no satisfactory solution to the problem of compiling linear algebra expressions with symbolic sizes.
We propose a solution for GMCs based on multi-versioning.
Multi-versioning~\cite{cooper1992procedure} is a compiler technique that entails generating different variants of the same function at compile-time and dynamically dispatching to the best-suited variant at run-time.
The technique is commonly used when vectorizing code (one variant with vectorization and another without), to resolve memory aliasing (one variant for non-overlapping arguments and another for overlapping arguments), and when optimizing a function for different microarchitectures. 
Multi-versioning has also been applied to domain-specific applications~\cite{lazcano2020runtime,jimborean2011handling} and to code generators that aim to better exploit task parallelism~\cite{thoman2014compiler}.
In the context of linear algebra, manual or semi-automatic multi-versioning is common practice when implementing low-level kernels such as \textsc{gemm} to cover different architectures and different parts of the input space. 

%% file: sections/2_gmc.tex
\section{Design of the Code Generator}\label{sec:gmc}

In contrast to standard matrix chains, the matrices in a GMC may be inverted and/or transposed and have features such as symmetry. 
While an MC is fully specified by the number and sizes of the matrices, the description of a GMC must contain more information. 
Specifically, a GMC with $n$ matrices entails $n - 1$ matrix multiplications, which we call \emph{associations}. 
Each matrix may be optionally transformed by \emph{unary operators}: transposition and inversion. 
Furthermore, each matrix carries certain \emph{features} (e.g., symmetric, invertible). 
The sequence of $n$ pairs of unary operators and matrix features defines the \emph{shape} of a chain.
The shape specifies all relevant aspects of a chain except for the matrix sizes. 
For example, consider the chain $G_1 L^{-1} U G_2^T$.
The matrix features are as follows: 
Both $G_1$ and $G_2$ are general matrices, $L$ is lower triangular and invertible, and $U$ is upper triangular.
There are no unary operators applied to $G_1$ and $U$, inversion is applied to $L$, and transposition to $G_2$. 
Collectively, this information specifies the shape of the chain.
We refer to a chain with unknown sizes as a \emph{symbolic chain}.

% introduce kernels for different parenthesizations, and 1 association => N kernels (e.g., symm, gemm)
%\p{If we follow the idea above, here we again draw the contrast with standard MC, where everything was GEMMs}
In an MC, each association is mapped to a matrix multiplication kernel (e.g., \textsc{gemm}). 
However, for a GMC, the code to generate for an association will depend both on the features and the unary operators of the two matrices involved.
For example, the association $G_1 G_2^T$ could be mapped to a call to \textsc{gemm}, whereas $L^{-1} G$ is better mapped to a call to \textsc{trsm}. 
We use the term \emph{kernel} to mean a function (callable from C/C\texttt{++}) that can perform associations for some set of combinations of features and unary operators. 
In this sense, both \textsc{gemm} and \textsc{trsm} are kernels.
Our code generator relies on a set of kernels that combined support all possible types of associations. 
%\lk{Refer to full list of kernels?}
%\p{see my next comment}
Note that there can be more than one kernel that supports a given association type. 
For example, if $S$ is symmetric, then $S G$ can be performed by either the kernel \textsc{symm} or \textsc{gemm}. 
Given that the choice of kernel for an association depends on the features of the operands, features matter also for the intermediate results. 
The code generator must therefore reason about how features propagate.

%\p{One aspect that is not currently highlighted: Given that the choice of kernel for an association depends on the properties of the left and right operands, properties matter not only for the input operands, but also for the intermediate ones. This means that when constructing a variant, we need to be able to reason about how properties propagate.}

% introduce parenthesization
There are many different sequences of associations that correctly evaluate a matrix chain of a given shape. 
A chain with $n$ matrices admits $C_{n-1} = \frac{(2n-2)!}{n!(n-1)!}$ distinct \emph{parenthesizations} ($C_{n}$ is the $n$-th Catalan number).
However, even for a fixed parenthesization, there may be several compatible sequences of associations.
This is, in part, because a parenthesization only partially orders the associations. 
For example, the parenthesization $(G_1 G_2)(G_3 G_4)$ can be sequenced with either $G_1 G_2$ or $G_3 G_4$ associated first. 
Another source of multiple options is that several kernels can support the same type of association.
For example, consider $G_1^{-1} G_2^{-1}$, which can be rewritten as $(G_2 G_1)^{-1}$. 
In the first form, we can use a kernel that explicitly inverts one matrix and then solves a linear system. 
In the second form, we can use a general matrix multiplication kernel and the inversion is propagated to the result. 
By assigning a compatible kernel to each association, we obtain a code \emph{variant}.
A variant can be directly translated to a C\texttt{++} function by the code generator. 
%\lk{remove the example of propagation and replace with the real source of ambiguity: multiple kernels supporting the same association. Use an example of inversion propagation to make this point. This example is also used later.}

% 1 chain => many parenthesizations => many variants / different kernels
In summary, a chain can have many parenthesizations and each parenthesization can be realized by multiple different code variants (sequences of kernel calls). 
Different variants may use different sets of kernels, even for the same parenthesization.

% overview
Our code generator (recall \figurename~\ref{fig:setup}) takes a symbolic chain as input (using the grammar in \figurename~\ref{fig:grammar_input}) and generates C\texttt{++} code that can efficiently evaluate any instance of the chain. 
In particular, the code generator outputs $k$ pairs of C\texttt{++} functions, each of which implements one of the variants and an associated cost function. 
The code generator also outputs a \emph{dispatch function} that the application calls to evaluate an instance of the chain.  
The dispatch function computes the cost of each variant and dispatches control to the variant with the least cost. 
%\p{I like this overview. It's probably a repetition of what comes in the Introduction, where we have to present and comment Fig.1. Still, I feel that it's very useful here to explain where we are and where we're going.}

\changed{
When we set the sizes of the matrices in a symbolic chain, we obtain an \emph{instance} of the chain.
The matrix sizes are specified by a one-dimensional vector $\boldsymbol{q} = (q_0, q_1, \ldots, q_n) \in \mathbb N^{n+1}$.
The number of instances is infinite.}

% The matrix sizes are specified by a tuple $\boldsymbol{q} := (q_0, q_1, \ldots, q_n) \in \mathbb N^{n+1}$.
% When we set the sizes of all matrices in a symbolic chain, we obtain an \emph{instance} of the chain. 
% The number of instances is infinite.

% sketch an overview
In Section~\ref{sec:gen_variants}, we present a deterministic procedure that the code generator uses to construct a specific variant for any given parenthesization. 
%The choice is made using heuristics that aim to find the most efficient variant.
%This is feasible because the parenthesization fixes the sizes of all intermediate matrices and the remaining choices are often clear. 
The code generator then needs to choose a subset of the parenthesizations for which to generate code variants. 
In Section~\ref{sec:theory}, we present novel theoretical results that the code generator uses to select a small set that offers provable performance guarantees. 
Specifically, performance will be within a constant factor from optimal on all instances. 
In Section~\ref{sec:expansion}, we describe a tunable procedure that gradually expands the selected parenthesizations to obtain increasingly good performance. 
This allows a user to balance the trade-off between performance and code size and dispatch run-time overhead.

\subsection{Matrix features}

%\paragraph{Features}
The features of a matrix are a combination of a \emph{structure} and a \emph{property}.
The structure reflects how the entries are arranged in memory.
We allow the following structures: \underline{G}eneral (i.e., a dense matrix), \underline{S}ymmetric, \underline{L}ower-triangular, and \underline{U}pper-triangular.
We use the symbols $G$, $S$, $L$, and $U$ to denote a matrix with the corresponding structure.
We use $M$ to denote a matrix with unspecified or unknown structure, and we use $X$ to denote intermediate results.
All structures, except the general structure, imply that the matrix is square.

The property determines whether a matrix is invertible and, if so, which kernels can solve a linear system with it as the coefficient matrix.
We allow the following properties: Singular, invertible, symmetric positive-definite, and orthogonal.

Some combinations of structure, property, and unary operators are invalid.
Others trigger a rewrite to a simpler form.
For example, the general structure cannot be combined with the symmetric positive-definite property (since the latter implies the symmetric structure), and the inversion unary operator cannot be applied to a matrix with the singular property.
A matrix whose features imply that the matrix is an identity matrix, such as any triangular structure combined with the orthogonal property, triggers a rewrite that removes it from the input expression.
Transposition is removed when applied to a matrix with the symmetric structure, and inversion is replaced by transposition when applied to a matrix with the orthogonal property.

\subsection{Parenthesizations and variants}

A parenthesization can be represented by an expression tree and partially orders the $n - 1$ associations.
A variant is constructed from a sequential ordering of the associations that is compatible with some parenthesization.
\changed{The $i$th association ($i = 1, 2, \ldots, n - 1$) in a variant combines, via some kernel $\textsc{k}_i$, an operand of size $q_{a_i} \times q_{b_i}$ with an operand of size $q_{b_i} \times q_{c_i}$, where $0 \leq a_i < b_i < c_i \leq n$.}
The resulting matrix has size $q_{a_i} \times q_{c_i}$ and the size symbol $q_{b_i}$ does not appear in any further association in the variant.
\changed{The variant is uniquely represented by the sequence ${\{(\textsc{k}_i, (a_i, b_i, c_i)) \}}_{i=1}^{n-1}$.}
% A variant can be represented by a sequence of index triplets ${\{ (a_i, b_i, c_i) \}}_{i=1}^{n-1}$, where $0 \leq a_i < b_i < c_i \leq n$ for all $i$.
For example, the variant that issues first the leftmost associations of the parenthesization $((M_1 M_2) M_3) (M_4 M_5)$ is represented by
\changed{
\begin{align*}
\{ 
(\textsc{k}_1, (0, 1, 2)), (\textsc{k}_2, (0, 2, 3)), \\ (\textsc{k}_3, (3, 4, 5)), (\textsc{k}_4, (0, 3, 5))
\}.
\end{align*}
}
Note that the association $M_1 M_2$ is issued before $M_4 M_5$.
\changed{Different parenthesizations have distinct sets of triplets $(a_i, b_i, c_i)$ and can also have differing sets of kernels $\textsc{k}_i$.}
%\changed{Note, as well, that triplets and, with almost complete certainty, kernels in the sequence change when a variant for a different parenthesization of the same chain is considered.}
For example, in the previous example, there is an alternative variant of the same parenthesization where $M_4 M_5$ is evaluated first.

\subsection{Cost functions}

The code generator aims to generate variants with minimal cost. % for certain instances.\franls{At Paolo and Lars: this is not correct, is it? Erase sentence?}
\changed{%The previously given representation of variants is particularly convenient when modeling the cost of kernel calls and variants.
Let $\phi_{\textsc{k}_i} : \mathbb{N}^3 \rightarrow \mathbb{R}$ be the cost function corresponding to the kernel $\textsc{k}_i$ used in the $i$th association of a variant $$A = \{(\textsc{k}_i, (a_i, b_i, c_i))\}_{i=1}^{n-1}.$$
We model the cost of $A$ on an instance $\boldsymbol{q} = (q_0, \ldots, q_n)$ by
\begin{displaymath}
    T(A, \boldsymbol{q}) = \sum_{(\textsc{k}, (a,b,c)) \in A} \phi_{\textsc{k}}(q_a, q_b, q_c), 
    %= \sum_{i=1}^{n-1} \phi_{\textsc{k}_i}(q_{a_i}, q_{b_i}, q_{c_i}),
\end{displaymath}
which depends not only on the triplets but also on the choice of kernels.
}

%% file: sections/4_algorithm_generation.tex
\section{From Parenthesization to Code Variant}\label{sec:gen_variants}

Recall that each parenthesization can be realized by many different variants (sequences of kernel calls).
The code generator uses heuristics to construct a single variant for each parenthesization. 
After this simplification, the code generator can focus on selecting a subset of parenthesizations. 
In the end, the chosen parenthesizations are translated to their corresponding variants and C\texttt{++} code is generated for those variants.
In this section, we describe how a variant is constructed from a given parenthesization. 

%We introduce and motivate the heuristics used along the way. 
%\lk{The following text has been pushed forward from the previous section. Might be used here instead.}
%\begin{itemize}
%    \item Through a number of algebraic transformations and rules
%that do not imply loss of generality (detailed in Sec 3), such
%as always issuing the leftmost available association first,
%a parenthesization is translated into a single variant. We
%denote the set of variants for a chain by $\mathcal A$.
%\end{itemize}

The partial ordering of associations implied by the parenthesization is extended to a total ordering by performing the left-most available association first. 
This results in a preliminary sequence of $n - 1$ associations. 
Then the following steps are performed on each association in order:
\begin{enumerate}
    \item \emph{Propagation of inversion.} Rewrite the association and propagate an inversion to the result (if at all).
    \item \emph{Kernel assignment.} Assign a compatible kernel to the association.
    \item \emph{Propagation of transposition.} Rewrite the association and propagate a transposition to the result (if at all).
    \item \emph{Inference of features and sizes.} Infer the features and sizes of the result.% from the features and sizes of the two operands.
%\lk{removed because it's not connected to the story}
%    \item \emph{Memory management.} Manage temporary memory for the operands and the result. 
\end{enumerate}
In the following, we detail each step.

\paragraph{Step 1: Propagation of inversion}
Associations where both operands are inverted, i.e., $M_1^{-1} M_2^{-1}$, may be computed by explicitly inverting both operands and multiplying them together, or by inverting one and solving a linear system with the other as the coefficient matrix.
However, since explicit inversions are undesirable due to numerical stability and performance, these associations are rewritten as $M_1^{-1} M_2^{-1} = {(M_2 M_1)}^{-1}$, which translates to a matrix multiplication and propagation of an inversion.
%The result of this association will eventually be an operand in some other association, where a linear system is solved or the inverse is again propagated.
%\p{It's easy to find examples where this transformation leads to more flops than needed. How is this addressed?}
%\franls{Please, let me know about these examples.}
%\lk{Paolo: $L^{-1} U^{-1}$: better not to propagate!}

%That is one type of association where the propagation of inversion takes place.
An association with just one inverted operand sometimes triggers a propagated inversion.
Consider the left-to-right parenthesization $X_2 := (L_1 G_2^{-1}) G_3$, where $L_1, G_2 \in \mathbb{R}^{m \times m}$ are non-singular and $G_3 \in \mathbb{R}^{m \times n}$.
The variant that performs
\begin{enumerate}
    \item $X_1 := L_1 G_2^{-1}$ by solving a general linear system with a triangular right-hand side using \textsc{getrsv}, with a cost of cost $8m^3/3$ FLOPs,
    \item $X_2 := X_1 G_3$ by multiplying two general matrices using \textsc{gemm}, with a cost of $2m^2 n$ FLOPs,
\end{enumerate}
has an overall cost of $8m^3/3 + 2m^2 n$ FLOPs.
The variant obtained after rewriting the first association, $(L_1 G_2^{-1}) G_3 = {(G_2 L_1^{-1})}^{-1} G_3$, performs
\begin{enumerate}
    \item $X_1 := G_2 L_1^{-1}$ by solving a triangular linear system with a general right-hand side via \textsc{trsm} with a cost of $m^3$ FLOPs,
    \item $X_2 := X_1^{-1} G_3$ by solving a general linear system with a general right-hand side via \textsc{gegesv}, with a cost of $2m^3/3 + 2m^2 n$ FLOPs,
\end{enumerate}
has an overall cost of $5m^3/3 + 2m^2 n$ FLOPs, which is always cheaper than the first variant.

%Thus, rewriting to propagate an inversion is sometimes desirable even if not necessary.
We propagate an inversion in these cases:
\begin{itemize}
    \item Both operands are inverted.
    \item One operand is inverted, the inverted operand is general or symmetric, and the non-inverted operand is orthogonal or non-singular triangular.
\end{itemize}
The second case is based on heuristics that aim to minimize the number of linear systems to solve with general or symmetric coefficient matrices, since those are more expensive than solving linear systems with a triangular coefficient matrix.

In the rare event that an inversion is propagated to the end result, an explicit inversion is forced.
%This can only happen when all matrices are square: Every matrix must be non-singular for the inversion to be propagated to the last result.
% In these cases, computing an explicit inverse will not drastically increase the cost of any variant, 
% \p{increase with respect to what? the sentence makes it sound as if the inversion could be avoided}
% since its cost is in the order $\mathcal O (n^3)$ and there will be multiple kernels with similar costs.

\begin{figure}
    \centering{}
    \includegraphics[width=\columnwidth]{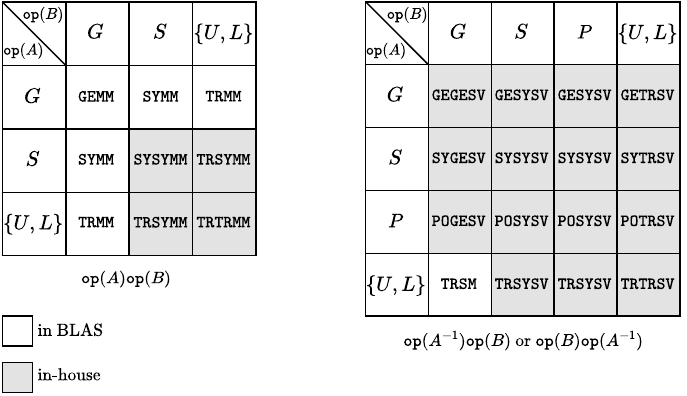}
    \caption{Mapping from features in the association to kernels for the product of matrices (left) and solving linear systems (right). In both tables, $\Op(X) = X, X^T$. Symmetric positive-definite matrices are denoted by $P$ on the right table. With a white background, kernels in BLAS.\@ With a gray background, kernels we have defined and implemented.
    }\label{fig:mapping_kernels}
\end{figure}

\paragraph{Step 2: Kernel assignment}
Each combination of features and unary operators in an association is mapped to a kernel.
Even though multiple kernels may be able to handle the same type of association, the code generator uses the best-fitting (most specialized) kernel for each type.
% Typically, one kernel is compatible with several types of associations.
\figurename~\ref{fig:mapping_kernels} presents the association-to-kernel mapping in the form of two lookup tables that are indexed by the features of the operands.
The left table is used when neither operand is inverted; the right table, when one is inverted.
The two operands in the association cannot be inverted at this step because of the rules for rewriting and propagation of inversion.

% \subsection{Propagate transpositions}
\paragraph{Step 3: Propagation of transposition}
Many kernels support implicitly transposed operands.
However, most kernels do not support all possible transposition patterns.
To avoid explicit transposition, propagating a transposition might be necessary to get an association with a transposition pattern supported by the assigned kernel.
If the transposition pattern is not supported by the assigned kernel, then a rewrite is performed and a transposition is propagated.
For example, if \textsc{trmm} is assigned to $L G^T$, then the association is rewritten to $(G L^T)^T$ and the outer transposition is propagated.
This is done because \textsc{trmm} does not support implicit transposition of its general operand.
This rewrite rule always results in a transposition pattern that is supported by the assigned kernel.
In the rare event that a transposition is propagated to the end result, an explicit transposition is performed.

\begin{figure}[htbp]
\centering
\includegraphics[width=\columnwidth]{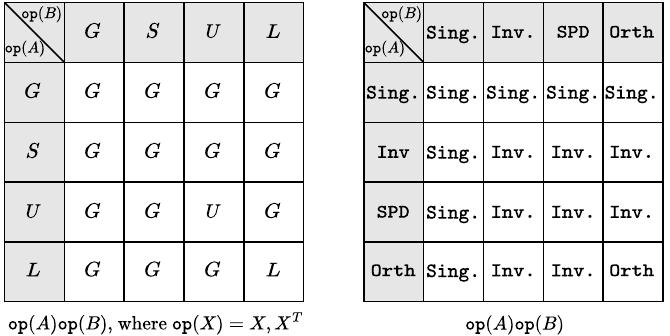}
\caption{Lookup tables for inference of structure (left) and property (right).
}\label{fig:features_LUT}
\end{figure}

\paragraph{Step 4: Inference of features and sizes}
To leverage specialized kernels, the code generator must infer the features of intermediate results.
Feature inference consists of two parts: Structure inference and property inference.
The rules are encoded as lookup tables, shown in \figurename~\ref{fig:features_LUT}.
The structure/property of the left operand identifies the row and the structure/property of the right operand identifies the column.
Note that the structure of a transposed triangular operand is the opposite triangular structure.
For example, in the association $X := U^T L$, the lower-triangular structure of $X$ is inferred from row $L$ (left operand) since $U^T$ has lower-triangular structure and column $L$ (right operand) in the table on the left.
The inference of features only takes into account the features of the operands without factoring in further algebraic relations that may be present.
For instance, in the association $Q^T G$, where $Q$ is orthogonal, the rules infer that the result has a general structure.
However, if $Q$ is the $Q$-factor from a QR-decomposition of $G$, then $Q^T G$ actually has an upper-triangular structure.
In these cases, a less efficient kernel might be eventually assigned, but it will never cause an error in the evaluation.

The symbolic sizes of all intermediate and final results are also inferred.
This allows the code generator to construct a symbolic cost function for each association and variant.

%% file: sections/5_theory.tex
\section{Theory for Variant Selection}\label{sec:theory}

The overhead of multi-versioning comes in two forms: Code size overhead due to the code generated for the selected variants, and run-time overhead due to the run-time variant selection and dispatch.
Since both overheads grow linearly with the number of generated variants, we aim to generate as few variants as possible. 
For matrix chains, selecting a good set of variants is crucial, since a bad choice leads to arbitrarily poor performance~\cite{lopez2025parenth}. 
For example, consider $G_1 G_2 G_3$ with its two parenthesizations $(G_1 G_2) G_3$ and $G_1 (G_2 G_3)$. 
The FLOP cost ratio of the latter to the former is $\frac{q_1 q_3 (q_0 + q_2)}{q_0 q_2 (q_1 + q_3)}$, which for \changed{instances of the form} $\boldsymbol{q} = (1, s, 1, s)$ has no upper bound as $s$ grows.
%Hence, generating code only for the latter parenthesization can result in computing arbitrarily many more operations on some instances. 
%\franls{at Paolo and Lars: do we need this example here? We already argue for this in the introduction of the paper.}
%\p{I see no problem in repeating an important fact. In case, we could do without defining $r(\boldsymbol{q})$. However, we need to remove or rephrase "leads to arbitrarily poor performance" since it appears twice in this paragraph.} 
%\franls{I touched the text above. Please double-check it.}
Since the previous section established a one-to-one mapping from parenthesizations to variants, we treat them as synonymous from now on.
In this section, given a chain with $n$ matrices, we show how to select a small set of at most $n + 1$ variants such that, for each instance, the best selected variant is within a constant factor from the optimal variant.
%\lk{Paolo makes the case that the care needed for variant selection is special and important and should be made already in the introduction.}

The theoretical results we present require kernel cost functions to be monotonically increasing in each argument.
Here, we only use FLOPs as the cost function.
Furthermore, for the sake of brevity, we do not consider transposition, as it does not impact the number of FLOPs of a variant.

Let $\mathcal A$ denote the set of all possible variants (one per parenthesization) using the construction in Section~\ref{sec:gen_variants}.
Let $\mathcal Z \subseteq \mathcal A$ denote the set of variants selected for code generation.
The optimal cost when restricted to variants in $\mathcal Z$ is, for any given instance $\boldsymbol{q}$, obviously greater than or equal to the optimal cost in the full set $\mathcal A$.
We introduce the notion of \emph{penalty (on an instance)} to quantify the relative increase in cost:
\begin{equation}
    \label{eq:penalty-function}
    P(\mathcal{Z}, \boldsymbol{q}) :=  
    \frac{
        \min_{
            Z \in \mathcal{Z}
        }
        ~T(Z, \boldsymbol{q})
    }{
        \min_{
            A \in \mathcal{A}
        }
        ~T(A, \boldsymbol{q})
    } 
    - 1.
\end{equation}
We let $P(\emptyset, \boldsymbol{q}) = \infty$ by convention.
For example, $P(\mathcal Z, \boldsymbol{q}) = 0.5$ means that on instance $\boldsymbol{q}$, by being restricted to variants in $\mathcal Z$, the optimal cost increases by $50\%$.
The penalty is zero only if the optimal cost in $\mathcal{Z}$ is the same as the optimal cost in $\mathcal A$.
We extend the penalty on an instance to measure the \textit{total penalty} over all instances:
\begin{equation}\label{eq:max-penalty-function}
    P(\mathcal{Z}) := \sup_{\boldsymbol{q} \in \mathbb{N}^{n+1}} P(\mathcal{Z}, \boldsymbol{q}).
\end{equation}
%As shown in~\cite{lopez2025parenth}, $P(\mathcal Z)$ may be unbounded even when $\mathcal Z$ is non-empty. \p{This is a bit of a repetition. We just showed an example of this fact in the opening of this section. We should move the reference [26] to the opening.}

We seek a small set of variants $\mathcal Z$ for which the total penalty $P(\mathcal Z) \leq \rho$ for some (ideally small) constant $\rho$.
A set of variants $\mathcal Z$ is called \emph{essential} if $P(\mathcal Z)$ is bounded and for all proper subsets $\mathcal S \subset \mathcal Z$, $P(\mathcal S)$ is unbounded.
For standard matrix chains it has been shown that there is a single well-defined essential set (of parenthesizations) of size $n + 1$~\cite{lopez2025parenth}.
\changed{
However, these results do not easily extend to generalized matrix chains for a number of reasons:
i) Variants for GMCs use a much broader set of kernels (as opposed to only \textsc{gemm}), each with a distinct cost function; 
ii) Variants for the GMC are built by propagating inverses to minimize the number of expensive linear systems to solve, which results in non-trivial kernel assignments;
iii) The kernels invoked often change from one variant to another.
For these reasons, a more complex mathematical machinery is needed to reach similar results for generalized matrix chains.
% In the remainder of this section we work our way towards similar results for generalized matrix chains.
}

% A kernel cost function maps a matrix size triplet $(a, b, c) \in \mathbb N_+^3$ to a cost. 
% \p{We're struggling between generality ("kernel cost function") and the specific costs we use (next sentence). It feels like we want to be general, but we're not.}
%Every cost function associated to some kernel we consider maps a matrix size triplet $(a, b, c) \in \mathbb N_+^3$ to some cost that depends on the sizes.
Each of the FLOP-based kernel cost functions that we use (see Table~\ref{tbl:kernel_costs} in Appendix~\ref{app:blas_extended}) belongs to one of the following types, where the coefficients are kernel-specific constants\footnote{For the sake of simplicity, we do not consider lower-order terms in the cost functions.}:
% Each kernel cost function (see Table~\ref{tbl:kernel_costs}) belongs to one of the following types, where the coefficients are kernel-specific constants\footnote{For the sake of simplicity, we do not consider lower-order terms in the cost functions of Type~I and Type~II.}:
\begin{itemize}
    \item Type~I:\@ $\phi(a,b,c) = \beta a b c$.
    \item Type~IIa:\@ $\phi(a,b,c) = \beta_1 a^3 + \beta_2 a^2 c$.
    \item Type~IIb:\@ $\phi(a,b,c) = \beta_1 c^3 + \beta_2 c^2 a$.
\end{itemize}
% \p{we do refer to Type II}
% \franls{Yes, what I meant is that we don't refer to Type~IIa and Type~IIb separately.}
% \lk{Fran, that's fine. The a and b here is purely to clarify the list. Had we had to make a fuzz about IIa being different from IIb below, then that's a sign that there are two different types (II and III) rather than two variants of the same type.}
% \franls{Okay, roger!}
Kernels that solve a linear system with a non-triangular coefficient matrix with a general rectangular right-hand side are Type~II.
Every other kernel is Type~I.

We start with Lemma~\ref{lemma:alpha_type_i_ii}, which establishes inequalities between Type~I and Type~II kernel cost functions.
%when the associations from which they stem are of some particular form.
%This result will be used in later proofs to guarantee the optimal cost in subsets of variants is always within a constant factor from the optimal cost in the full set of variants.

%\p{I don't see why such a pair should always exist. It doesn not, right? -- notice that the kernel is fixed ($k_i$)}
% \franls{The lemma only says: ``Assume there is some term $t_{\rm e}$ of this form and there is some term $t_{\rm o}$ of that form. Then, there is a constant such that $t_{\rm e} \leq \alpha t_{\rm o}$.''. Here we don't question whether these terms always exist or not. We simply prove that a bound exists between the two terms when they share a pair of consecutive indices, regardless of whether they are Type~I or Type~II.}

%\p{again, we use mix general and specific notions. We use $\phi$, but then Type I \& II.}
% \franls{I have removed the Type~I and Type~II specifications. Even so, I don't think $\phi$ is more general than it needs to be.}

%\lk{At Fran: The statement of the lemma is too general. The proof relies on Type I and Type II, which implies that we only consider our specific set of cost functions. The statement currently does not make that restriction and thus they may be of any type}

\begin{lemma}\label{lemma:alpha_type_i_ii}
    Let $\boldsymbol{q} = (q_0, q_1, \ldots, q_n)$ be an instance and let $m$ be an index such that $q_m = \min_i q_i$.
    If $t_{\rm e}$ is a term of the form $\phi_{\textsc{k}_{\rm e}}(q_{j-1}, q_j, q_m)$ or $\phi_{\textsc{k}_{\rm e}}(q_m, q_{j-1}, q_j)$ in the cost function of one variant, and $t_{\rm o}$ is a term of the form $\phi_{\textsc{k}_{\rm o}}(q_{j-1}, q_j, q_z)$ or $\phi_{\textsc{k}_{\rm o}}(q_z, q_{j-1}, q_j)$ in the cost function of a variant (possibly the same), then there exists some constant $\alpha \in \mathbb{R}_+$ such that $t_{\rm e} \leq \alpha t_{\rm o}$.
\end{lemma}
\begin{IEEEproof}
  The pair of adjacent sizes $(q_{j-1}, q_j)$ in the triplets implies that both terms come from the association of $M_j \in \mathbb{R}^{q_{j-1} \times q_j}$ with another matrix.
  Let $A$ denote the matrix whose association with $M_j$ produces the term $t_{\rm e}$.
  The term $t_{\rm e}$ is the cost of either $\operatorname{op}(M_j) \operatorname{op}(A)$, yielding $t_{\rm e} = \phi_{\textsc{k}_{\rm e}}(q_{j-1}, q_j, q_m)$, or $\operatorname{op}(A) \operatorname{op}(M_j)$, yielding $t_{\rm e} = \phi_{\textsc{k}_{\rm e}}(q_m, q_{j-1}, q_j)$.
  The size of $A$ is $q_j \times q_m$ in the former case and $q_m \times q_{j-1}$ in the latter.
  Similarly, let $B$ denote the matrix whose association with $M_j$ produces $t_{\rm o}$.
  The size of $B$ is $q_j \times q_z$ when $t_{\rm o} = \phi_{\textsc{k}_{\rm o}}(q_{j-1}, q_j, q_z)$, and $q_{z} \times q_{j-1}$ when $t_{\rm o} = \phi_{\textsc{k}_{\rm o}}(q_z, q_{j-1}, q_j)$.
  We show that independently of whether $t_{\rm e}$ and $t_{\rm o}$ are a Type~I or Type~II kernel cost, there exists some constant $\alpha \in \mathbb{R}_{+}$ for which the inequality $t_{\rm e} \leq \alpha t_{\rm o}$ holds.
    We only cover one case here, the other cases are similar and can be found in Appendix~\ref{app:proof}.
    
    \textbf{Case I}: Both $t_{\rm{e}}$ and $t_{\rm{o}}$ are Type~I.
    Since $q_m \leq q_z$, we have
    \begin{displaymath}
        t_{\rm{e}} = \beta_1 q_{j-1} q_j q_m \leq \beta_1 q_{j-1} q_j q_z = \frac{\beta_1}{\beta_2} \beta_2 q_{j-1} q_j q_z = \alpha t_{\rm o},
    \end{displaymath}
    where $\alpha := \frac{\beta_1}{\beta_2}$ is a constant.
\end{IEEEproof}

Lemma~\ref{lemma:alpha_type_i_ii} can be extended to also cover the final association in the variants, where $t_{\rm e} = \phi_{\textsc{k}_{\rm e}}(q_0, q_m, q_n)$ and $t_{\rm o} = \phi_{\textsc{k}_{\rm o}}(q_0, q_z, q_n).$

In what follows, variants of a particular kind---which are called \emph{fanning-out} variants and were first introduced by López et al.~\cite{lopez2025parenth}---play a central role.
For each $h \in \{0,1,\ldots,n\}$, there is a fanning-out variant $E^h \in \mathcal{A}$ defined by the parenthesization
\begin{equation}\label{eq:both-ways}
    (M_1 ( \cdots (M_{h-1} M_h) \cdots ))(( \cdots (M_{h+1} M_{h+2})\cdots ) M_n),
\end{equation}
where (i) the prefix $M_1 \cdots M_h$ is computed right-to-left, (ii) the suffix $M_{h+1} \cdots M_n$ is computed left-to-right, and (iii) once prefix and suffix have been computed, the two partial results are associated.
If $h \in \{ 0, n \}$, then either the prefix or the suffix is the entire chain and two of the three steps vanish.
We denote the set of fanning-out variants by $\mathcal{E} := \{E^0, E^1, \ldots, E^n\} \subseteq \mathcal{A}$.
The number of fanning-out variants is $n-1$ when $n \leq 3$, and $n+1$, otherwise.

The next lemma shows that, for any given instance, there is some fanning-out variant whose cost is within a constant factor from the optimal cost.

\begin{lemma}\label{lemma:bound_Em}
  Let $\boldsymbol{q} = (q_0, q_1, \ldots, q_n)$ be any instance and $m$ be any index such that $q_m = \min_i q_i$.
  Let $T_{\rm opt}$ be the optimal cost on $\boldsymbol{q}$.
  Then, there exists a constant $\hat{\alpha}$ such that $T(E^{m}, \boldsymbol{q}) < 2 \hat{\alpha} T_{\rm opt}$.
\end{lemma}
\begin{IEEEproof}
  The case $n \leq 2$ is trivial, so assume $n \geq 3$.
  The core of the proof is a matching between terms in the cost functions of the variants such that pairs of terms of the forms considered in Lemma~\ref{lemma:alpha_type_i_ii} are established and, thus, a bound can be set.

  % Each matrix $M_j$ is consumed by some association in every variant.\p{"in every variant"?!} \franls{Yes}
  % This association gives rise to a term with the pair of sizes with consecutive indices $(q_{j-1}, q_j)$ as argument.% in every variant's cost function.
  % Such a pair appears as arguments in only one term per variant cost function, which has the form $\phi_{\textsc{k}_i}(q_{j-1}, q_j, q_z)$ or $\phi_{\textsc{k}_i}(q_z, q_{j-1}, q_j)$.
  % The pair $(q_0, q_n)$ appears as argument to the last term in every variant's cost function.
  % A maximum of two consecutive pairs can be present in the same term: When two adjacent matrices are associated, as in $M_j M_{j+1}$, yielding the term $\phi_{\textsc{k}_i}(q_{j-1}, q_j, q_{j+1})$ with two adjacent pairs.
  % \p{it feels like we're going through baby steps here. Given the 10-page limit, we might give ideas for the proof, and move more quickly} 

  For each variant, there is exactly one association that involves the matrix $M_j$.
  This association produces a cost term where the adjacent pair of sizes $(q_{j-1}, q_j)$ appears in the argument list.
  This term either has the form $\phi_{\textsc{k}_{i}}(q_{j-1}, q_j, q_z)$ or $\phi_{\textsc{k}_{i}}(q_z, q_{j-1}, q_j)$.
  %Said pair only appears in one term per variant cost function.
  The pair $(q_0, q_n)$ appears in exactly one term in each variant cost function and is produced by the final association.
  At most two consecutive pairs of sizes can appear in the same term, namely, when the association is of the form $M_j M_{j+1}$ and hence the term is of the form $\phi_{\textsc{k}_{i}}(q_{j-1}, q_j, q_{j+1})$.

  The cost function $T_E := T(E^{m}, \boldsymbol{q})$ has the form
  \begin{align*}
      T_{E} &= \underbrace{\phi_{\textsc{k}_{1}}(q_{m-2}, q_{m-1}, q_m) + \cdots + \phi_{\textsc{k}_{m-1}}(q_0, q_1, q_m)}_{\textrm{prefix}}\\ 
      &+ \underbrace{\phi_{\textsc{k}_{m}}(q_m, q_{m+1}, q_{m+2}) + \cdots + \phi_{\textsc{k}_{n-2}}(q_m, q_{n-1}, q_n)}_{\textrm{suffix}} + \gamma,
  \end{align*}
  where $\gamma = \phi_{\textsc{k}_{n-1}}(q_0, q_m, q_n)$ if $m \neq \{0, n\}$; otherwise, $\gamma$ is not present.
  Let $\sigma$ be the function that for each $j \in \{1, \ldots, n\} \setminus \{m, m+1\}$ maps a term $t_{\rm e}$ in $T_E$ of the form
  \begin{align*}
      \phi_{\textsc{k}_i}(q_{j-1}, q_{j}, q_m) \quad &\text{if}\quad j < m,\\
      \phi_{\textsc{k}_i}(q_m, q_{j-1}, q_{j}) \quad &\text{if}\quad j > m+1, \nonumber
  \end{align*}
  to a term $t_{\rm o}$ of the form $\phi_{\hat{\textsc{k}}_i}(q_{j-1}, q_{j}, q_z)$ or $\phi_{\hat{\textsc{k}}_i}(q_z, q_{j-1}, q_{j})$ in $T_{\rm{opt}}$, and the term $\phi_{\textsc{k}_{n-1}}(q_0, q_m, q_n)$, if it exists, in $T_E$ to the term $\phi_{\hat{\textsc{k}}_{n-1}}(q_0, q_z, q_n)$ in $T_{\rm{opt}}$.
  
  The function $\sigma$ is well-defined. 
  Only one term in $T_E$ and $T_{\rm{opt}}$ contains a given pair of adjacent sizes, and every term in $T_E$ contains an adjacent pair, except for $\gamma$, which, if it exists, is mapped separately.
  
  The terms in $T_{\rm{opt}}$ are matched with none, one, or two terms in $T_E$.
  Since $2 T_{\rm{opt}}$ has two copies of each term and $\sigma$ cannot map more than two terms in $T_E$ to the same term in $T_{\rm{opt}}$, each term $t_{\rm e}$ in $T_E$ can be paired with a unique term $t_{\rm o}$ in $2 T_{\rm{opt}}$.
  By Lemma~\ref{lemma:alpha_type_i_ii}, $t_{\rm e} \leq \alpha t_{\rm o}$ holds for some constant $\alpha$.
  
  Let $\hat \alpha$ be the largest $\alpha$ across all terms in $T_E$. 
  %Let $\psi(t_{\rm e}, \sigma(t_{\rm e}))$ be a function that maps every $t_{\rm e}$ and the $t_{\rm o}$ it is mapped to via $\sigma$ to an $\alpha$ that makes the inequality given by Lemma~\ref{lemma:alpha_type_i_ii} hold.
  %Let $\hat{\alpha} = \sup_{t_{\rm e} \in T_E} \psi(t_{\rm e}, \sigma(t_{\rm e}))$.
  Since $T_E$ has strictly fewer terms than $2 T_{\rm opt}$ and all terms in $2 T_{\rm{opt}}$ are positive, $T_E < 2 \hat{\alpha} T_{\rm opt}$.
\end{IEEEproof}

%Give examples of \alpha for different chains. Show how \alpha varies depending on the features of the operands.
When Lemma~\ref{lemma:bound_Em} is applied to the standard matrix chain, the only kernel is \textsc{gemm}.
One can verify, after establishing the mapping of terms and calculating the different $\alpha$'s, that $\hat{\alpha} = 1$, yielding a bound of $T_E < 2 T_{\rm opt}$ as previously shown~\cite{lopez2025parenth}.
By contrast, if the chain is of the form $G_1 \cdots G_{i-1} L_i G_{i+1} \cdots G_n$, the kernels \textsc{gemm} and \textsc{trmm} are used, and one can verify that $\hat{\alpha} = 2$, yielding $T_E < 4 T_{\rm opt}$.
In general, the value of $\hat{\alpha}$ is bounded above by $8$, yielding  $T_E < 16 T_{\rm opt}$.
This can be verified by computing $\alpha$ according to Lemma~\ref{lemma:alpha_type_i_ii} for every possible pair of kernels in Table~\ref{tbl:kernel_costs} in Appendix~\ref{app:blas_extended}.

The next theorem shows that by selecting only the fanning-out variants $\mathcal E$, we can ensure that the generated code cannot have arbitrarily poor performance on any instance, since its total penalty is finite.

\begin{theorem}\label{thm:bound_E}
    The set of fanning-out variants $\mathcal E$ has finite total penalty.
\end{theorem}
\begin{IEEEproof}
    We show that there exists some constant $\rho$ such that $P(\mathcal E) \leq \rho$.
    Let $\boldsymbol{q} = (q_0, q_1, \ldots, q_n)$ be any instance and let $m$ be any index such that $q_m = \min_i q_i$.
    Since $E^m \in \mathcal{E}$, we can apply Lemma~\ref{lemma:bound_Em} to conclude that
    \begin{displaymath}
        T(E^m, \boldsymbol{q}) \leq 2 \hat \alpha_{\boldsymbol{q}} \min_{A \in \mathcal A} T(A, \boldsymbol{q})
    \end{displaymath}
    for some constant $\hat \alpha_{\boldsymbol{q}}$. 
    Since $\hat \alpha_{\boldsymbol{q}} \leq 8$, 
    \begin{displaymath}
        P(\mathcal{E}) = \sup_{\boldsymbol{q} \in \mathbb{N}^{n+1}} P(\mathcal{E}, \boldsymbol{q}) = \sup_{\boldsymbol{q} \in \mathbb{N}^{n+1}} \frac{\min_{E \in \mathcal{E}} T(E, \boldsymbol{q})}{\min_{A \in \mathcal{A}} T(A, \boldsymbol{q})} - 1 \leq 15. %2 \hat{\alpha}_* - 1.
    \end{displaymath}
    Therefore, we can take $\rho = 15$.
\end{IEEEproof}
Note that the constant $\rho = 15$ is in general very pessimistic.

If a matrix $M_i$ in the chain is necessarily square (e.g., it is symmetric or inverted), then $q_{i-1}$ and $q_i$ are bound by equality.
We denote this equivalence relation by $q_{i-1} \sim q_i$, which partitions the size symbols into equivalence classes.
We let $C[q_i]$ denote the equivalence class of $q_i$. 

There are $n_{\rm c} = n - n_{\rm sq} + 1$ equivalence classes, where $n_{\rm sq}$ is the number of square matrices in the symbolic chain. 
%Let $n_{\rm sq}$ be the number of necessarily square matrices.
%Then there are $n_{\rm fs} = n - n_{\rm sq} + 1$ free sizes.
%Each free size is associated with a unique equivalence class $C[q_i]$.
%Let each equivalence class be represented by the size with the least index.
%Let $\hat{\boldsymbol{q}}$ of length $n_{\rm fs}$ be a reduced instance vector containing only the representative sizes. 
For example, the equivalence classes for the chain $S_1 G_2 S_3 L_4 G_5$ are 
$    \{ q_0, q_1 \}, 
    \{ q_2, q_3, q_4 \},
    \{ q_5 \}.
$
%and the reduced instance vector is $\hat{\boldsymbol{q}} = (q_0, q_2, q_5)$.

The next theorem shows how to construct a subset of $\mathcal{E}$ that also has finite total penalty.

\begin{theorem}\label{thm:E_s}
    Assume $n \geq 4$.
    Let $\mathcal E_{\rm s}$ be a set of $n_{\rm c}$ variants constructed as follows. 
    For each equivalence class $C$, select a size variable $q_h \in C$ and add the corresponding fanning-out variant $E^h$ to $\mathcal E_{\rm s}$.
    Then, $P(\mathcal E_{\rm s})$ is finite.

    %Partition $\{ 0, 1, \ldots, n \} = \mathcal C_1 \cup \cdots \cup \mathcal C_{n_{\rm fs}}$ such that $i, j \in \mathcal C_k$ only if $C[q_i] = C[q_j]$.
    %Define for $k = 1, 2, \ldots, n_{\rm fs}$ the subsets of fanning-out variants $\mathcal E_k := \{ E^h\ |\ h \in \mathcal C_k \}$.
    %Let $\mathcal E_{\rm s}$ be any set of variants obtained by selecting exactly one variant from each $\mathcal E_k$.
    %Then the penalty of $\mathcal E_{\rm s}$ is finite.
\end{theorem}
\begin{IEEEproof}
    Let $\boldsymbol{q} = (q_0, q_1, \ldots, q_n)$ be any instance, and let $m$ be any index such that $q_m = \min_i q_i$. 
    Lemma~\ref{lemma:bound_Em} ensures that $E^m$ is within a constant factor from optimal on $\boldsymbol{q}$.
    Let $q_{z} \in C[q_m]$ be any other size in the same equivalence class as $q_m$.
    Since $q_{z} = q_m$, Lemma~\ref{lemma:bound_Em} also applies to $E^{z}$ and thus $E^{z}$ is also within a constant factor from optimal on $\boldsymbol{q}$.
    %In general, all variants in $\mathcal E_k$ are within a constant factor from optimal due to Lemma~\ref{lemma:bound_Em}, when $q_m \in C_k$. 
    %Therefore, selecting one variant from each part suffices to guarantee finite penalty on all instances.
    %Since $\mathcal E_{\rm s}$ has one variant from each $\mathcal E_k$, its penalty is finite.
    Since $\mathcal E_{\rm s}$ has one variant from each equivalence class, regardless of which size is minimal on any given instance, the set contains a variant whose cost is never arbitrarily far from the optimal cost.
\end{IEEEproof}

In summary, we have reduced the exponentially many variants for the GMCP to a set $\mathcal E_{\rm s}$ with at most $n+1$ variants with finite total penalty (i.e., the best-in-set is never too far from the optimal cost).

%% file: sections/6_expansion.tex
\section{Empirically Expanding a Set of Variants}\label{sec:expansion}

Sets of variants generated as per Theorem~\ref{thm:E_s} are never arbitrarily far from optimal in terms of FLOPs.
However, the total penalty, especially when execution time is considered, may still be too large for a given application.
In this section, we introduce an empirical procedure to expand sets of variants to balance the trade-off between performance and overhead.

\changed{
We first informally present the problem.
Assume a set of variants has been produced as per Theorem~\ref{thm:E_s} whose performance is deemed unsatisfactory.
In this setting, one cannot replace any variant in the produced set, since the bound on the penalty would be broken.
Hence, one can only resort to adding more variants to the set, effectively increasing the overhead in terms of both code size and run-time dispatch.
%We assume the maximum size of the resulting set is an input.
The problem is to select additional variants to add to the original set such that the performance of the resulting set is maximized while keeping the total number of variants below a fixed threshold.
}

\changed{We now give a formal definition of the problem.}
As before, let $\mathcal A$ denote the set of all variants for a chain.
Let $\mathcal Z_0 \subset \mathcal A$ be an initial (possibly empty) subset, $K$ a non-negative integer \changed{representing the maximum number of selected variants}, and $F: 2^{\mathcal A} \rightarrow \mathbb{R}$ an objective function that assigns a score to each \changed{possible} subset \changed{of variants} (lower is better).
The goal is to find a set $\mathcal Z \supseteq \mathcal Z_0$, subject to the constraint $| \mathcal Z | \leq K$, that minimizes $F(\mathcal Z)$.
%This is a cardinality-constrained combinatorial optimization problem.

\changed{
%Intuitively, since the problem is formulated as a minimization, $F$ must be a function that captures the inverse of the performance of sets of variants.
%Note that the performance of a set of variants is subject to the instances upon which it should be evaluated.
%This is, if no instances are considered, there is no sound way to rank two distinct sets of variants of the same cardinality: $F$ is not well-defined\footnote{Trivially, if two sets are considered and one is a superset of the other, there is a clear best in terms of performance.}.
Since the objective $F(\mathcal Z)$ depends on an infinite number of instances, we must finitely sample the instance space to obtain a computable objective function. 
%Hence, finding an optimal solution to the problem is, except for particular cases (such as $K = |\mathcal{A}|$), intractable, so we settle for an empirical approach.
Given a set of $\ell$ sampled instances $\mathcal{Q} = \{\boldsymbol{q}_1, \ldots, \boldsymbol{q}_{\ell}\}$, we consider sampled objective functions $F : 2^{\mathcal{A}} \times \mathcal{Q} \rightarrow \mathbb{R}$ that assign a score to each possible set of variants based on the per-instance penalties on $\mathcal{Q}$.
Examples of such objective functions include the \emph{maximum penalty} and the \emph{average penalty}:
\begin{displaymath}
    F_{\rm max}(\mathcal Z, \mathcal Q) = \max_{i=1}^\ell p_i, \quad\quad F_{\rm avg}(\mathcal Z, \mathcal Q) = \frac{1}{\ell} \sum_{i=1}^\ell p_i,
\end{displaymath}
where $p_i := P(\mathcal Z, \boldsymbol{q}_i)$ is the penalty of $\mathcal Z$ on instance $\boldsymbol{q}_i$.
%Since smaller penalties translate to a better performance and so should smaller values of $F$, we require objective functions to be monotonically increasing in the penalties of the input set on the instances $\boldsymbol{q}_i$.

We present in Algorithm~\ref{alg:greedy_seccp} a greedy algorithm that in each iteration adds the variant that decreases the value of $F$ the most. 
%Since increasing the size of a given set can only lower its penalties on $\mathcal{Q}$ and, hence, diminish the value of $F$, such algorithm will reach the cardinality constraint without ever increasing the value of the objective function.
}

\changed{
\begin{algorithm}[htbp]
    \caption{$\mathcal{Z} = \mathtt{ExpandSet}(\mathcal{A}, \mathcal{Q}, F, K, \mathcal{Z}_0)$}\label{alg:greedy_seccp}
    \begin{flushleft}
        \textbf{Input:} \\
        \hspace{1em} $\mathcal{A}$, the set of all variants for a shape; \\
        \hspace{1em} $\mathcal{Q} \subset \mathbb N^{n+1}$, a set of $\ell$ sampled instances; \\
        \hspace{1em} $F : 2^{\mathcal A} \times \mathcal Q \rightarrow \mathbb R$, an objective function to minimize; \\
        \hspace{1em} $K \in \mathbb N$, the maximum cardinality of the expanded set; \\
        \hspace{1em} $\mathcal{Z}_0 \subseteq \mathcal A$, an initial, possibly empty, set of variants; \\

        \textbf{Output:} \\
        \hspace{1em} $\mathcal{Z} \subseteq \mathcal A$, the expanded set of variants with $| \mathcal Z | \leq K$
    \end{flushleft}
    \hrule
    \begin{algorithmic}[1]
        \State $\mathcal{Z} \gets \mathcal{Z}_0$\;
        \State $v_{\rm min} \gets F\left( \mathcal Z, \mathcal Q \right)$ if $\mathcal{Z} \neq \emptyset$, otherwise $v_{\rm min} \gets \infty$\; 
        \While{$|\mathcal{Z}| < K$} 
            \State $C \gets \emptyset$\;  \Comment{Initialize candidate to empty}
            \State $v_{*} \gets \infty$\; 
            \For{$D \in \mathcal{A} \setminus \mathcal{Z}$} \Comment{For every variant not in $\mathcal{Z}$}
                \State $\hat{\mathcal{Z}} \gets \mathcal{Z} \cup \{D\}$\; \Comment{Create prospective superset}
                \If{$F(\hat{\mathcal{Z}}, \mathcal{Q}) < v_{*}$}
                    \State $v_{*} \gets F(\hat{\mathcal{Z}}, \mathcal{Q})$\; \Comment{Update best value}
                    \State $C \gets D$\; \Comment{Update best candidate}
                \EndIf{}
            \EndFor
            % \State Let $D \in \mathcal{A} \setminus \mathcal{Z}$ be a variant that minimizes $F_{\mathcal{Q}}\left(\mathcal{Z} \cup \{D\} \right)$\;
            % \State $v_* \gets F_{\mathcal Q}\left( \mathcal{Z} \cup \{D\} \right)$\;
            \If{$v_* \geq v_{\rm min}$}
                \State \Return $\mathcal{Z}$\; \Comment{Return if value of $F$ is not improved}
            \EndIf{}
            \State $\mathcal{Z} \gets \mathcal{Z} \cup \{ C \}$\; \Comment{Add variant that most improves $F$}
            \State $v_{\rm min} \gets v_*$\; \Comment{Update value of $F(\mathcal{Z}, \mathcal{Q})$}
        \EndWhile
        \State \Return $\mathcal{Z}$\;
    \end{algorithmic}
\end{algorithm}
}

%% file: sections/7_experiments.tex
\section{Experiments}\label{sec:experiments}

%\franls{Note to self: carefully read this section -- present tense!!}
%\lk{Change to past tense when describing things we already did.}

We performed two experiments.
\changed{
The first one focuses on FLOPs and assesses both how far the performance of the generated code is from optimal and the effectiveness of the expanding procedure (Section~\ref{sec:expansion}).
The second experiment focuses on execution time, compares different flavors of generated code (including Armadillo as a point of reference), shows the effectiveness of the expanding procedure, and tests whether the expansion is improved when based on performance models instead of FLOPs.
}
% old introductory paragraph
% With the first one, we aim to assess the quality of the variant set $\mathcal E_{\rm s}$ (Theorem~\ref{thm:E_s}) with FLOPs as the cost function, and the effectiveness of the expanding procedure (Section~\ref{sec:expansion}).
% The second experiment aims to assess the quality of $\mathcal E_{\rm s}$ with execution time as the cost function, the effectiveness of the expanding procedure, and whether or not the set expansion improves when based on performance models instead of FLOPs.

\subsection{Experiment with FLOPs}\label{subsec:exp1}

% n_sqm (number sq matrices) // n_sqc (number sq combinations)
% for n=7
% n_sqm | #shapes
% 0     |    1
% 1     |    63
% 2     |    1701
% 3     |    25515
% 4     |    229635
% 5     |    1240029
% 6     |    3720087
% total = 5217031 = 10^7 - 9^7
% for each n_sqm = choose(n, n_sqm) * (n_sqc ** n_sqm)

\changed{
We measured the deviation from optimal of generated sets across many different instances and shapes for various chain lengths.

For each shape, we constructed a set of variants $\mathcal{E}_{\rm s}$, as per Theorem~\ref{thm:E_s}, which minimized the average penalty over a training set with $10^5$ random instances with sizes in the range $2 \leq q_i \leq 1000$.
From $\mathcal{E}_{\rm s}$, we performed one and two steps of Algorithm~\ref{alg:greedy_seccp}, forming the expanded sets $\mathcal{E}_{\rm s1}$ and $\mathcal{E}_{\rm s2}$, respectively.
As a point of reference, we consider the left-to-right order our compiler generates, denoted by $\mathcal{L}$.
Note that this left-to-right evaluation order is more advanced than what some programming languages (e.g., MATLAB) normally do, since it is built by inferring properties of intermediate operands, propagating transposition and inversion operators, and leveraging a broad set of kernels.
This is further evidenced in the second experiment, where we compare against Armadillo.
We computed the ratio of the number of FLOPs of the best variant in each generated set to the number of FLOPs of the overall optimal variant on a validation set with $10^3$ random instances per shape taken from the same distribution as the training set.

To render the experiment feasible, we restricted features to 10 options per matrix: 
No transpositions; singular or inverted general matrices; symmetric positive-definite matrices possibly inverted; upper-/lower-triangular matrices possibly non-singular and possibly inverted.
We further limited the lengths to $n = 5,6,7$ and required at least one matrix per chain to be rectangular.
Since 9 of the 10 options imply that the matrix is square, the base set $\mathcal E_{\rm s}$ contains only two or three variants in more than $95\%$ of the shapes.
The set never contains fewer than two variants, since at least one matrix is rectangular.
In total, for each $n$, we tested all the $10^n - 9^n$ different shapes that meet these requirements and, for each shape, we tested $10^3$ instances.
}

% OLD PARAGRAPH.
% For each shape, we constructed a set of variants $\mathcal{E}_{\rm s}$, as per Theorem~\ref{thm:E_s}, which minimized the respective metric over a training set with $10^5$ random instances with sizes in the range $2 \leq q_i \leq 1000$.
% From $\mathcal E_{\rm s}$, we performed one and two steps of Algorithm~\ref{alg:greedy_seccp}, thus forming the expanded sets $\mathcal E_{\rm s1}$ and $\mathcal E_{\rm s2}$, respectively.
% As a point of reference, we consider the left-to-right order, denoted by $\mathcal{L}$, which is customary in various programming languages (e.g., MATLAB).
% We computed the metrics of the four sets of variants on a validation set with $10^4$ random instances taken from the same distribution as the training set.

\begin{figure*}[htbp]
  \centering
  \includegraphics[width=\textwidth]{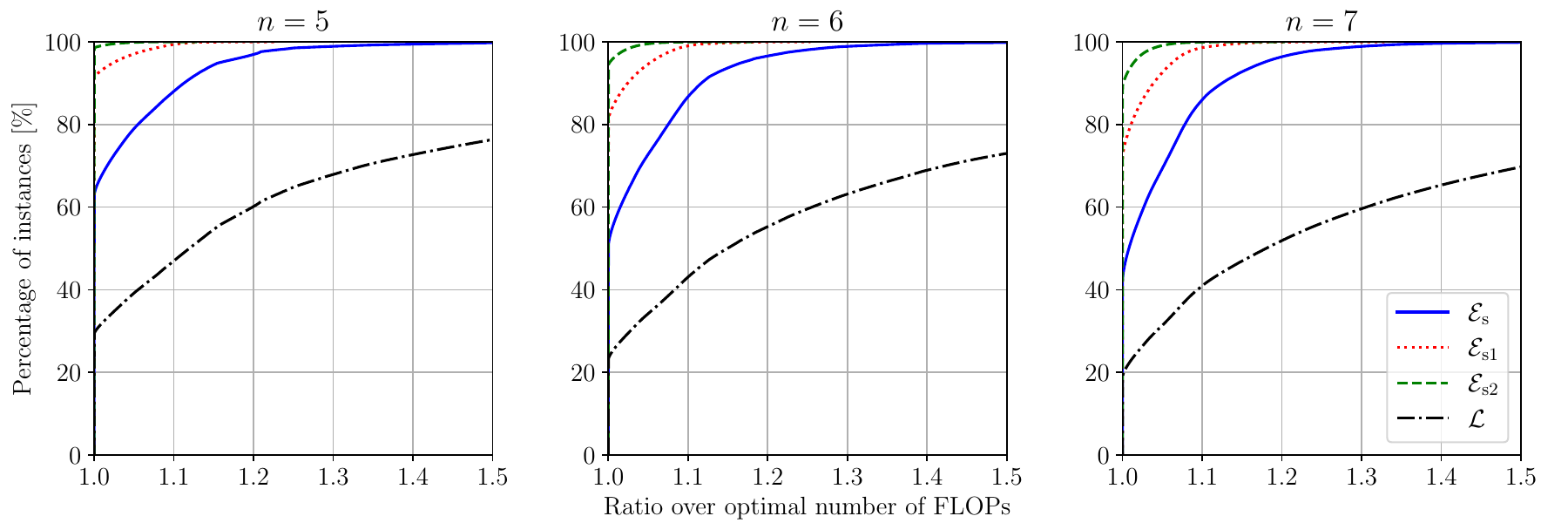}
  \caption{Empirical cumulative distribution functions of the ratio over optimum on a per-instance basis, measured on FLOPs, of the base sets $\mathcal E_{\rm s}$ (blue solid line), the sets after expanding by one (red dotted line) and two (green dashed line), and the singleton with the left-to-right variant (black dash-dotted line), for $n=5,6,7$. For a given set of variants $\mathcal{S}$, and a given point $x_0$ on the x-axis, the corresponding value $y_0$ on the y-axis indicates the percentage of instances for which the best variant in $\mathcal{S}$ computes at most $x_0$ times more FLOPs than the optimum.
  }\label{fig:exp1}
\end{figure*}

%The number of shapes for which $\mathcal{E}_{\rm{s}} = n + 1 - n_{\rm{sq}}$ is exactly ${n \choose n_{\rm{sq}}} 9^{n_{\rm{sq}}}$.
%\franls{I strongly dislike the level of specificity in the next paragraph. Should we instead give an average size of $\mathcal{E}_{\rm{s}}$? Table?}
%For $n=5$, $| \mathcal{E}_{\rm{s}} | = 2,3,4,5,6$, for $32805$, $7290$, $810$, $45$, and $1$ chain, respectively.
%For $n=6$, $| \mathcal{E}_{\rm{s}} | = 2,3,4,5,6,7$, for $354294$, $98415$, $14580$, $1215$, $54$, and $1$ chain, respectively.
%For $n=7$, $| \mathcal{E}_{\rm{s}} | = 2,3,4,5,6,7,8$, for $3720087$, $1240029$, $229635$, $25515$, $1701$, $63$, and $1$ chain, respectively.

\changed{
\figurename~\ref{fig:exp1} shows the empirical cumulative density function (eCDF) of the ratio over optimum of the distinct sets of variants across all test instances for each $n$.
The graphs reveal the percentage of instances (y-axis; 0\% to 100\%) for which the ratio was at or below some given value (x-axis).
For example, with $n = 5$, the left-to-right order had a ratio over optimum at or below $1.2$ (x-axis) on 60\% (y-axis) of the instances.
Since the ratio over optimum of the left-to-right variant reaches values above 465 for all $n$, the plots do not show the full range on the x-axis.

% max ratio computed on raw data for each set: [Es, Es1, Es2, LtR].
% max_values n=5: [  1.7357631   1.2345679   1.1482903 465.60004  ]
% max_values n=6: [  1.9546742   1.310606    1.2684586 552.8148   ]
% max_values n=7: [  2.1160226   1.6125      1.375     556.1492   ]

For the left-to-right variant $\mathcal{L}$, the ratio over optimum is above 465 on some instances for every $n$.
For this set, we observed a ratio above 1.5 on more than 23\% of the instances for all $n$.
The theory says that, for most shapes, the total penalty and, thus, the ratio over optimum, of $\mathcal{L}$ is unbounded.
This is why $\mathcal{L}$ exhibits such large ratios over optimum for some instances.
This demonstrates the potential risk of a compiler for expressions with symbolic sizes that generates code only for the left-to-right evaluation (or any other single parenthesization), even if some optimizations are applied while crafting the variant.

For the base set, $\mathcal{E}_{\rm s}$, the ratio over optimum is below $2.1$ for all instances and $n$.
Also, this ratio is at or below 1.2 on 96\% of the instances across all $n$.
Hence, the set given by theory at worst performs roughly twice the optimal number of FLOPs on some instance and, on the vast majority of instances, it performs less than 20\% more FLOPs than the optimal.

For the two expanded sets, $\mathcal{E}_{\rm s1}$ and $\mathcal{E}_{\rm s2}$, the largest observed ratios over optimum were 1.62 and 1.38, respectively, both for $n=7$.
The ratio was at or below 1.05 for more than $92$\% and $99$\% of the instances across all $n$, respectively.
That is, expanding by just one or two variants makes the increase over optimum negligible (5\% more FLOPs than optimum) on most instances.
Hence, the expanding procedure proves to be an effective way to improve performance with respect to the base set.
The base set, in turn, outperforms the left-to-right variant, especially with regards to setting a tight upper bound on the deviation from optimum.
This confirms that a compiler for symbolic sizes must generate code for more than just one variant if performance guarantees are desired.
}

\subsection{Experiment with execution time}\label{subsec:exp2}
% \franls{Need to reintroduce?}
% With the second experiment we assess the quality of $\mathcal E_{\rm s}$ from Theorem~\ref{thm:E_s} with execution time as the cost function, assess the effectiveness of the expansion procedure (Section~\ref{sec:expansion}), and evaluate the impact of selecting variants based on performance models instead of FLOPs.

% experimental setup
\changed{
This experiment focuses on the execution time of the generated code.
Time measurements were taken on an Intel Xeon Gold 6132 processor nominally running at 2.60~GHz with 192~GB of memory.
The code was compiled with the GCC C\texttt{++} compiler version $13.3.0$ with the flags \texttt{-O3 -march=native} and linked to multi-threaded OpenBLAS version 0.3.27.
Here, Armadillo $14.6.1$ is used as a reference point.
Armadillo was linked to the same OpenBLAS version and guaranteed to have access to its run-time library.
We used all $14$ physical cores in the processor and pinned threads to cores.
Each time measurement was repeated ten times and summarized by the median.
}

We constructed performance models by timing each kernel on a 3D/2D/1D Cartesian grid with six points per axis over the range $[50,1000]$ ($50, 100, 300, 500, 700, 1000$).
For each point, we recorded the performance (FLOP/s).
To estimate the execution time of a kernel call, the corresponding model estimates the performance by interpolating the grid samples.
The FLOP count is then divided by the estimated performance to obtain the execution time.
The execution time of a variant is estimated by summing the estimates for the variant's kernel calls.

% methodology
We fixed $n=7$ and randomly sampled $10^3$ shapes with the same ten options per matrix as in the first experiment.
Each matrix in a chain was given a $50\%$ probability of being rectangular.
The other nine options were given equal probability.
We also required at least one rectangular matrix per chain.

\begin{figure}[htbp]
  \centering
  \includegraphics[width=\columnwidth]{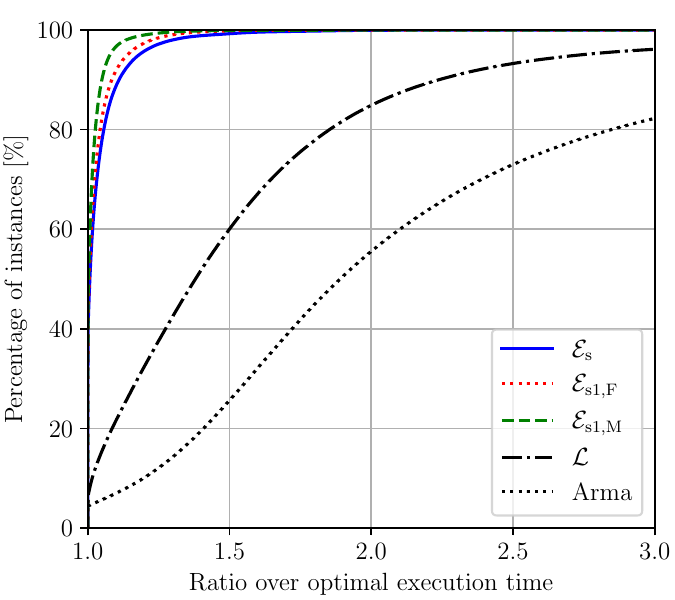}
  \caption{Empirical cumulative distribution functions of the ratio of the execution time of various flavors of generated code, the left-to-right evaluation, and Armadillo to the optimal execution time for $10^6$ random instances (with $n=7$).
  The flavors of generated code are the base theory-infused set (blue solid line), the sets after increasing the sets by $1$ variant using FLOPs (red dotted line) and performance models (green dashed line). 
  The singleton with the left-to-right variant (black dash-dotted line) and Armadillo (black dotted line) are used as references.
  }\label{fig:exp2_1}
\end{figure}

\changed{
For each sampled shape, the following was done.
A set of variants $\mathcal{E}_{\rm{s}}$ was constructed as per Theorem~\ref{thm:E_s} to minimize the average penalty based on FLOPs over a training set with $10^5$ random instances with sizes in the range $50 \leq q_i \leq 1000$.
The set $\mathcal{E}_{\rm{s}}$ was then expanded via Algorithm~\ref{alg:greedy_seccp} to increase its size by one:
Once using FLOPs, producing $\mathcal E_{\rm s1,F}$, and once using performance models, producing $\mathcal E_{\rm s1,M}$.
The left-to-right variant $\mathcal L$ was again used as an in-house point of reference.
For each shape, we also generated Armadillo code that exploits as much knowledge of the input matrices as possible\footnote{We used \texttt{symmatl}, \texttt{trimatl}, and \texttt{trimatu} to specify matrix properties, and \texttt{inv\_sympd} to specify the inversion operator on an SPD matrix.}.
We computed the ratio of the execution time of different flavors of generated code ($\mathcal{E}_{\rm s}$, $\mathcal{E}_{\rm{s1,F}}$, $\mathcal{E}_{\rm{s1,M}}$), the left-to-right variant ($\mathcal{L}$), and Armadillo to the execution time of the optimal variant on a validation set with $10^3$ random instances.
In total, we test on $10^6$ instances.

\figurename~\ref{fig:exp2_1} shows the eCDF of the ratio over the optimal execution time of the generated code and Armadillo for all the test instances across all shapes.

For $\mathcal{E}_{\rm{s}}$, the largest ratio over optimum was $9.24$ for some instance.
For $\mathcal{E}_{\rm{s1,F}}$ and $\mathcal{E}_{\rm{s1,M}}$, the largest ratio over optimal was $6.64$ and $7.43$, respectively.
For $\mathcal{L}$ and Armadillo, the largest ratios were $128.74$ and $46.34$, respectively.
The percentage of instances with a ratio over optimum below $1.1$ was $96.7\%$, $91.9\%$ $88.8\%$, $21.6\%$, and $7.0\%$, respectively for $\mathcal{E}_{\rm{s1,M}}$, $\mathcal{E}_{\rm{s1,F}}$, $\mathcal{E}_{\rm{s}}$, $\mathcal{L}$, and Armadillo.
The percentage of instances with a ratio over optimum above $1.5$ was approximately $0.1\%$, $0.2\%$, $0.7\%$, $40.0\%$, and $74.4\%$, respectively for $\mathcal{E}_{\rm{s1,M}}$, $\mathcal{E}_{\rm{s1,F}}$, $\mathcal{E}_{\rm{s}}$, $\mathcal{L}$, and Armadillo.

In conclusion, the left-to-right variant performs poorly, as expected, although in general terms it outperforms Armadillo.
This is not completely surprising, given that our left-to-right evaluation infers features of intermediate operands, is able to propagate operators (avoiding solving expensive linear systems), and can use more specialized kernels than Armadillo.
The left-to-right variant is clearly outperformed by the theory-infused set, which performs well not only in terms of FLOPs (Section~\ref{subsec:exp1}) but also in terms of execution time.
Moreover, expanding the base set with rather simple performance models proves to be more effective that merely using FLOPs.
The average speed-ups of $\mathcal{E}_{\rm s}$, $\mathcal{E}_{\rm s1,F}$, and $\mathcal{E}_{\rm s1,M}$ over Armadillo across all test instances are, respectively, $2.30$, $2.32$, and $2.34$.
For some instances, the speed-ups of $\mathcal{E}_{\rm s}$, $\mathcal{E}_{\rm s1,F}$, and $\mathcal{E}_{\rm s1,M}$ over Armadillo reach values above $44.1$.
}

%% file: sections/8_conclusions.tex
\section{Conclusions}\label{sec:conclusion} % and future work?

We addressed the problem of compiling generalized matrix chains when the sizes of the matrices are not known at compile-time.
We presented and evaluated a code generator based on multi-versioning. 
%In this paper, we have presented a generator that produces efficient code for generalized matrix chains with symbolic sizes when FLOPs and execution time are considered.
The code generator relies on new theoretical results that help select a small set of code variants whose performance cannot be arbitrarily far from optimal.
The generator augments the theory with an empirical expansion procedure that further reduces the performance gap in practice.
In this way, one can balance the trade-off between code size overhead and performance of the generated code.

Experiments with FLOPs in Section~\ref{subsec:exp1} provided evidence that the sets given by the theoretical results here presented (Theorem~\ref{thm:E_s}) are often close to optimal.
By contrast, the left-to-right order, commonly used by state-of-the-art libraries and languages, performed much worse on average and in extreme cases was observed to require more than $500$ times more operations than optimal.
(Similarly poor performance is expected of any other single variant.)
Experiments also showed that the empirical expansion procedure had a significant effect even after adding just one or two variants. 

Experiments with execution time in Section~\ref{subsec:exp2} provided evidence that similar conclusions hold when measuring execution time instead of FLOPs.
In the same experiment, the expansion procedure was shown to provide better results when using performance models (even crude ones) instead of only FLOP counts.
\changed{
We included Armadillo as a point of reference, which was outperformed by the left-to-right evaluation our code generator produced.
However, it should be noted that Armadillo was not explicitly designed to handle generalized matrix chains and only chains with up to 4 matrices are considered\footnote{This was confirmed in a personal communication with Conrad Sanderson and Ryan Curtin, the main developers behind Armadillo.}.}

The present work does not consider common subexpressions and is restricted to generalized matrix chains.
The Common Subexpression Elimination optimization does not straightforwardly extend to matrices: If and how to apply such optimization is an NP-complete problem.
Considering more general expressions involving addition and subtraction adds further complications such as factoring out common operands.

In summary, we have presented a code generator that embodies the first theoretically sound and practical approach to the problem of compiling generalized matrix chains with symbolic sizes.
This work is one step towards a linear algebra compiler for general expressions with symbolic sizes, which so far remains an unsolved problem.

%\p{Conclusions are strong. Good to go.}

%% file: sections/appendix_proof.tex
\subsection{Lemma~\ref{lemma:alpha_type_i_ii}}\label{app:proof}

The following is the complete proof for Lemma~\ref{lemma:alpha_type_i_ii}.
% \lk{The prefix of the Lemma 1 proof is in the main text. It's probably slightly different than the version here. Moreover, we do not have to repeat the prefix in the interest of keeping the paper short. We can therefore simply continue where the prefix ended, i.e., resume here with Case II.}
% \lk{If the full proof will still be given here, then the common prefix has to be exactly the same as in the main text.}
% \p{Agreed}
% \franls{The prefix is the same in both places now. Feel free to erase the prefix here.}
\begin{IEEEproof}
  The pair of adjacent sizes $(q_{j-1}, q_j)$ in the triplets implies that both terms come from the association of $M_j \in \mathbb{R}^{q_{j-1} \times q_j}$ with another matrix.
  Let $A$ denote the matrix whose association with $M_j$ produces the term $t_{\rm e}$.
  The term $t_{\rm e}$ is the cost of either $\operatorname{op}(M_j) \operatorname{op}(A)$, yielding $t_{\rm e} = \phi_{\textsc{k}_{\rm e}}(q_{j-1}, q_j, q_m)$, or $\operatorname{op}(A) \operatorname{op}(M_j)$, yielding $t_{\rm e} = \phi_{\textsc{k}_{\rm e}}(q_m, q_{j-1}, q_j)$.
  The size of $A$ is $q_j \times q_m$ in the former case and $q_m \times q_{j-1}$ in the latter.
  Similarly, let $B$ denote the matrix whose association with $M_j$ produces $t_{\rm o}$.
  The size of $B$ is $q_j \times q_z$ when $t_{\rm o} = \phi_{\textsc{k}_{\rm o}}(q_{j-1}, q_j, q_z)$, and $q_{z} \times q_{j-1}$ when $t_{\rm o} = \phi_{\textsc{k}_{\rm o}}(q_z, q_{j-1}, q_j)$.
  We show that independently of whether $t_{\rm e}$ and $t_{\rm o}$ are a Type~I or Type~II kernel cost, there exists some constant $\alpha \in \mathbb{R}_{+}$ for which the inequality $t_{\rm e} \leq \alpha t_{\rm o}$ holds.
  Each case is handled separately.   
  
  \textbf{Case I}: Both $t_{\rm{e}}$ and $t_{\rm{o}}$ are Type~I.
  Since $q_m \leq q_z$, it is
  \begin{displaymath}
      t_{\rm{e}} = \beta_1 q_{j-1} q_j q_m \leq \beta_1 q_{j-1} q_j q_z = \underbrace{\tfrac{\beta_1}{\beta_2}}_{\alpha} \underbrace{\beta_2 q_{j-1} q_j q_z}_{t_{\rm{o}}}.
  \end{displaymath}

  \textbf{Case II}: $t_{\rm{e}}$ is Type~I and $t_{\rm{o}}$ is Type~II.\@
  Since all Type~II kernels solve a linear system of equations with a general right-hand-side, either the first two or the last two arguments in the cost function are bound by equality.
  There are three subcases:
  \begin{itemize}
      \item $t_{\rm{o}}$ comes from $M_j^{-1} B$ or $B M_j^{-1}$ and $t_{\rm{e}}$ comes from $M_j$ being consumed by a matrix product following a propagated inversion or by solving a linear system with a right-hand-side with a structure that forces the matrix to be square (if the matrix were not necessarily square, the kernel would be Type~II instead of Type~I).
      Therefore, $q_m \sim q_{j-1} \sim q_j$.
      It is
      \begin{align*}
      t_{\rm{e}} = \beta_1 q_m^3 &= \frac{\beta_1}{\beta_2 + \beta_3} (\beta_2 + \beta_3) q_m^3 \\
      &= \frac{\beta_1}{\beta_2 + \beta_3}(\beta_2 q_m^3 + \beta_3 q_m^3) \\
      &\leq \underbrace{\tfrac{\beta_1}{\beta_2 + \beta_3}}_{\alpha} (\underbrace{\beta_2 q_m^3 + \beta_3 q_m^2 q_z}_{t_{\rm{o}}}).
      \end{align*}

      \item $t_{\rm{o}}$ comes from $M_j B^{-1}$ (implying $q_j \sim q_z$) and $t_{\rm{e}}$ comes from either $A M_j$ or $M_j A$.
      It is
      \begin{align*}
      t_{\rm{e}} &= \beta_1 q_{j-1} q_j q_m \leq \beta_1 q_j^2 q_{j-1} = \frac{\beta_1}{\beta_3} (\beta_3 q_j^2 q_{j-1})\\ 
      &\leq \underbrace{\tfrac{\beta_1}{\beta_3}}_{\alpha} (\underbrace{\beta_2 q_j^3 + \beta_3 q_j^2 q_{j-1}}_{t_{\rm{o}}}).    
      \end{align*}
      
      \item $t_{\rm{o}}$ comes from $B^{-1} M_j$ (implying $q_z \sim q_{j-1}$) and $t_{\rm{e}}$ comes from either $A M_j$ or $M_j A$.
      It is
      \begin{align*}
      t_{\rm{e}} = \beta_1 q_{j-1} q_j q_m &\leq \tfrac{\beta_1}{\beta_3} (\beta_3 q_{j-1}^2 q_j)\\
       &\leq \underbrace{\tfrac{\beta_1}{\beta_3}}_{\alpha} (\underbrace{\beta_2 q_{j-1}^3 + \beta_3 q_{j-1}^2 q_j}_{t_{\rm{o}}}).
      \end{align*}
  \end{itemize}

  \textbf{Case III}: $t_{\rm{e}}$ is Type~II and $t_{\rm{o}}$ is Type~I.\@
  There are three subcases:
  \begin{itemize}
      \item $t_{\rm{e}}$ comes from $M_j^{-1} A$ or $A M_j^{-1}$ and $t_{\rm{o}}$ comes from associating $M_j$ either through propagation of the inversion and computation of a matrix product or solving a linear system with a right-hand-side whose sizes are bound by equality.
      Therefore, $q_z \sim q_{j-1} \sim q_j$.
      Theus
      \begin{displaymath}
      t_{\rm{e}} = \beta_1 q_j^3 + \beta_2 q_j^2 q_m \leq \beta_1 q_j^3 + \beta_2 q_j^3 = \underbrace{\tfrac{\beta_1 + \beta_2}{\beta_3}}_{\alpha} \underbrace{\beta_3 q_j^3}_{t_{\rm{o}}}.
      \end{displaymath}

      \item $t_{\rm{e}}$ comes from $M_j A^{-1}$ (implying $q_j \sim q_m$) and $t_{\rm{o}}$ comes from either $B M_j$ or $M_j B$.
      Thus
      \begin{align*}
      t_{\rm{e}} = \beta_1 q_m^3 + \beta_2 q_m^2 q_{j-1} &\leq \beta_1 q_{j-1} q_j q_z + \beta_2 q_{j-1} q_j q_z\\
      &= \underbrace{\tfrac{\beta_1 + \beta_2}{\beta_3}}_{\alpha} \underbrace{\beta_3 q_{j-1} q_j q_z}_{t_{\rm{o}}}.
      \end{align*}

      \item $t_{\rm{e}}$ comes from $A^{-1} M_j$ (implying $q_m \sim q_{j-1}$) and $t_{\rm{o}}$ comes from either $B M_j$ or $M_j B$.
      Thus 
      \begin{align*}
      t_{\rm{e}} = \beta_1 q_m^3 + \beta_2 q_m^2 q_j &\leq \beta_1 q_{j-1} q_j q_z + \beta_2 q_{j-1} q_j q_z\\
      &= \underbrace{\tfrac{\beta_1 + \beta_2}{\beta_3}}_{\alpha} \underbrace{\beta_3 q_{j-1} q_j q_z}_{t_{\rm{o}}}.
      \end{align*}
  \end{itemize}

  \textbf{Case IV}: Both $t_{\rm{e}}$ and $t_{\rm{o}}$ are Type~II.\@
  %Since they belong to type (ii), no more than two arguments in each cost function can be bound by equality.
  There are five subcases:
  \begin{itemize}
      \item $t_{\rm{e}}$ comes from either $M_j^{-1} A$ or $A M_j^{-1}$ and $t_{\rm{o}}$ comes from either $M_j^{-1} B$ or $B M_j^{-1}$.
      In either case, $q_{j-1} \sim q_j$ and it holds
      \begin{align*}
      t_{\rm{e}} 
      &= \beta_1 q_j^3 + \beta_2 q_j^2 q_m 
      = \tfrac{\beta_1}{\beta_3} \beta_3 q_j^3 + \tfrac{\beta_2}{\beta_4} \beta_4 q_j^2 q_m \\
      &\leq (\tfrac{\beta_1}{\beta_3} + \tfrac{\beta_2}{\beta_4}) (\beta_3 q_j^3 + \beta_4 q_j^2 q_m)\\
      &\leq (\underbrace{\tfrac{\beta_1}{\beta_3} + \tfrac{\beta_2}{\beta_4}}_{\alpha}) (\underbrace{\beta_3 q_j^3 + \beta_4 q_j^2 q_z}_{t_{\rm{o}}}).
      \end{align*}

      \item $t_{\rm{e}}$ comes from $A^{-1} M_j$ and $t_{\rm{o}}$ comes from $B^{-1} M_j$.
      Therefore, $q_m \sim q_z \sim q_{j-1}$.
      It holds 
      \begin{align*}
      t_{\rm{e}} = \beta_1 q_m^3 + \beta_2 q_m^2 q_j &= \tfrac{\beta_1}{\beta_3} \beta_3 q_m^3 + \tfrac{\beta_2}{\beta_4} \beta_4 q_m^2 q_j\\
       &\leq (\underbrace{\tfrac{\beta_1}{\beta_3} + \tfrac{\beta_2}{\beta_4}}_{\alpha}) (\underbrace{\beta_3 q_m^3 + \beta_4 q_m^2 q_j}_{t_{\rm{o}}}).
      \end{align*}
      
      \item $t_{\rm{e}}$ comes from $M_j A^{-1}$ and $t_{\rm{o}}$ comes from $M_j B^{-1}$.
      Therefore, $q_m \sim q_z \sim q_j$.
      It holds
      \begin{align*}
      t_{\rm{e}} &= \beta_1 q_m^3 + \beta_2 q_m^2 q_{j-1} = \tfrac{\beta_1}{\beta_3} \beta_3 q_m^3 + \tfrac{\beta_2}{\beta_4} \beta_4 q_m^2 q_{j-1}\\
      &\leq (\underbrace{\tfrac{\beta_1}{\beta_3} + \tfrac{\beta_2}{\beta_4}}_{\alpha}) (\underbrace{\beta_3 q_m^3 + \beta_4 q_m^2 q_{j-1}}_{t_{\rm{o}}}).
      \end{align*}
      
      \item $t_{\rm{e}}$ comes from $A^{-1} M_j$ and $t_{\rm{o}}$ comes from $M_j B^{-1}$.
      Therefore, $q_m \sim q_{j-1}$ and $q_j \sim q_z$.
      It holds
      \begin{align*}
      t_{\rm{e}} 
      &= \beta_1 q_m^3 + \beta_2 q_m^2 q_z 
      = \tfrac{\beta_1}{\beta_3} \beta_3 q_m^3 + \tfrac{\beta_2}{\beta_4} \beta_4 q_m^2 q_z \\
      &\leq (\tfrac{\beta_1}{\beta_3} + \tfrac{\beta_2}{\beta_4}) (\beta_3 q_m^3 + \beta_4 q_m^2 q_z) \\
      &\leq (\underbrace{\tfrac{\beta_1}{\beta_3} + \tfrac{\beta_2}{\beta_4}}_{\alpha}) (\underbrace{\beta_3 q_z^3 + \beta_4 q_z^2 q_m}_{t_{\rm{o}}}).
      \end{align*}

      \item $t_{\rm{e}}$ comes from $M_j A^{-1}$ and $t_{\rm{o}}$ comes from $B^{-1} M_j$.
      Therefore, $q_j \sim q_m$ and $q_z \sim q_{j-1}$.
      We have 
      \begin{align*}
      t_{\rm{e}} 
      &= \beta_1 q_m^3 + \beta_2 q_m^2 q_z 
      = \tfrac{\beta_1}{\beta_3} \beta_3 q_m^3 + \tfrac{\beta_2}{\beta_4} \beta_4 q_m^2 q_z \\
      &\leq (\tfrac{\beta_1}{\beta_3} + \tfrac{\beta_2}{\beta_4}) (\beta_3 q_m^3 + \beta_4 q_m^2 q_z) \\
      &\leq (\underbrace{\tfrac{\beta_1}{\beta_3} + \tfrac{\beta_2}{\beta_4}}_{\alpha}) (\underbrace{\beta_3 q_z^3 + \beta_4 q_z^2 q_m}_{t_{\rm{o}}}).    
      \end{align*}
  \end{itemize}

  The same line of reasoning can be used to show that a term of the form $\phi_{\boldsymbol{k}_i} (q_0, q_m, q_n)$, with $m \neq \{0, n\}$, is a constant factor away from a term of the form $\phi_{\boldsymbol{k}_i} (q_0, q_z, q_n)$, with $z \neq \{0, n\}$.
  For the sake of brevity, we do not specify all the cases here.
\end{IEEEproof}

%% file: sections/appendix_kernels.tex
\subsection{Kernels}\label{app:blas_extended}

%We present a list with kernels to complement the existing functionality in BLAS and LAPACK.\@

In order to cover all possible kinds of associations, we must complement the functionality in BLAS and LAPACK with custom kernels. 
The kernels are divided into two classes; the naming convention is similar to BLAS and LAPACK.
The first class contains kernels for computing specific matrix products;
the names of the kernels follow the format \textsc{xxmm} or \textsc{xxyymm}.
The second class contains kernels for solving linear systems;
the names follow the format \textsc{xxsv} or \textsc{xxyysv}.
Typically, names with four letters (e.g., \textsc{trmm}) denote kernels that associate a general matrix with a matrix of the structure or property specified by the first two letters, whereas names with six letters (e.g., \textsc{potrsv}) are given to kernels that associate two non-general matrices.
There are some kernels in the second class whose names we have elongated to avoid ambiguity with existing kernels in LAPACK, such as \textsc{gegesv}, as opposed to \textsc{gesv}, which already exists but partially supports the functionality we needed for the paper.
When the kernel solves a linear system, the first two letters denote the features of the coefficient matrix and the following two letters, the features of the right-hand side.

In this appendix, inverses are explicitly denoted and $\Op(X) = X, X^T$.
Furthermore, $P$ is used to denote a symmetric positive-definite matrix (\emph{not} a permutation matrix).
And $L$ is used to denote a lower- or upper-triangular matrix.
We use $m$ to denote the number of rows of the leftmost matrix, $k$ for the number of columns of the leftmost matrix, and $n$ for the number of columns of the rightmost matrix.
If one input matrix to a kernel is necessarily square, then we only use $m$ and $n$.
If both input matrices are necessarily square, we only use $m$.

\begin{table*}[htbp]
  \centering{}
  \caption{The kernels, their cost functions, and their assigned associations.}\label{tbl:kernel_costs}
    \begin{tabular}{m{1.3cm} m{6cm}  m{6cm}  m{2cm} } 
     \toprule
     Kernel & Computation & Cost function $\phi(m, n, k)$ (FLOPs) & Associations \\ 
     \midrule
     \textsc{gemm} & $C := \alpha * \Op(A) * \Op(B) + \beta * C$ & $2 m k n$ & $\Op(G_1) \Op(G_2)$ \\ 
     \midrule
     \textsc{symm} & $C := \alpha * A * \Op(B) + \beta * C $,  where $A$ is symmetric & $2 m^2 n$ if $A$ is on the left; otherwise, $2 m n^2$ & $S \Op(G)$, $\Op(G) S$\\
     \midrule
     \textsc{trmm} & $B := \alpha * \Op(A) * B$ or $B := \alpha * B * \Op(A)$, where $A$ is triangular & $m^2 n$ if $A$ is on the left; otherwise, $m n^2$ & $\Op(L) \Op(G)$, $\Op(G)\Op(L)$\\
     \midrule
     \textsc{sysymm} & $C := \alpha * A * B + \beta C$, where $A,B$ are symmetric & $2 m^3$ & $S_1 S_2$ \\
     \midrule
     \textsc{trsymm} & $B := \alpha * \Op(A) * B$ or $B:= \alpha * B * \Op(A)$, where $A$ is triangular and $B$ is symmetric & $m^3$ & $\Op(L) S$, $S \Op(L)$ \\
     \midrule
     \textsc{trtrmm} & $C:=\alpha * \Op(A) * \Op(B)$, where $A,B$ are triangular matrices & $m^3/3$ if $\Op(A)$ and $\Op(B)$ have the same triangularity; otherwise, $\frac{2}{3}m^3$ & $\Op(L_1)\Op(L_2)$ \\
     \midrule
     \textsc{gegesv} & Solve $\Op(A) * X = B$ or $X * \Op(A) = B$, where $A$ and $B$ are general matrices. $B$ is overwritten by the solution matrix X & $\frac{2}{3}m^3 + 2 m^2 n$ if $A$ is on the left; otherwise, $\frac{2}{3}n^3 + 2n^2 m$ & $\Op(G_1^{-1})\Op(G_2)$, $\Op(G_2) \Op(G_1^{-1})$  \\
     \midrule
     \textsc{gesysv} & Solve $\Op(A) * X = B$ or $X * \Op(A) = B$, where $A$ is general and $B$ is symmetric. $B$ is overwritten by the solution matrix X & $\frac{8}{3}m^3$ & $\Op(G^{-1}) S$, $S \Op(G^{-1})$  \\
     \midrule
     \textsc{getrsv} & Solve $\Op(A) * X = B$ or $X * \Op(A) = B$, where $A$ is general and $B$ is triangular. $B$ is overwritten by the solution matrix X & $2 m^3$ if $A$ is on the left and $B$ is lower-triangular or if $A$ is on the right and $B$ is upper-triangular; otherwise, $\frac{8}{3}m^3$  & $\Op(G^{-1})\Op(L)$, $\Op(L) \Op(G^{-1})$  \\
     \midrule
     \textsc{sygesv} & Solve $A * X = B$ or $X * A = B$, where $A$ is symmetric and $B$ is general. $B$ is overwritten by the solution matrix X & $m^3 / 3 + 2 m^2 n$ if $A$ is on the left; otherwise, $n^3 / 3 + 2 mn^2$ & $S^{-1} \Op(G)$, $\Op(G) S^{-1}$  \\
     \midrule
     \textsc{sysysv} & Solve $A * X = B$ or $X * A = B$, where $A$ and $B$ are symmetric. $B$ is overwritten by the solution matrix X & $\frac{7}{3}m^3$ & \makecell[l]{$S_1^{-1} S_2$, \\ $S_2 S_1^{-1}$}  \\
     \midrule
     \textsc{sytrsv} & Solve $A * X = B$ or $X * A = B$, where $A$ is symmetric and $B$ is triangular. $B$ is overwritten by the solution matrix X & $\frac{7}{3}m^3$ & $S^{-1} \Op(L)$, $\Op(L) S^{-1}$ \\
     \midrule
     \textsc{pogesv} & Solve $A * X = B$ or $X * A = B$, where $A$ is symmetric positive-definite and $B$ is general. $B$ is overwritten by the solution matrix X & $m^3/3 + 2 m^2 n$ if $A$ is on the left; otherwise, $n^3/3 + 2 m n^2$ & $P^{-1} \Op(G)$, $\Op(G) P^{-1}$ \\
     \midrule
     \textsc{posysv} & Solve $A * X = B$ or $X * A = B$, where $A$ is symmetric positive-definite and $B$ is general. $B$ is overwritten by the solution matrix X & $\frac{7}{3}m^3$ & \makecell[l]{$P^{-1} S$, \\ $S P^{-1}$} \\
     \midrule
     \textsc{potrsv} & Solve $A * X = B$ or $X * A = B$, where $A$ is symmetric positive-definite and $B$ is triangular. $B$ is overwritten by the solution matrix X & $\frac{5}{3}m^3$ if $A$ is on the left and $B$ is lower-triangular or if $A$ is on the right and $B$ is upper-triangular; otherwise, $\frac{7}{3}m^3$ & $P^{-1}\Op(L)$, $\Op(L) P^{-1}$ \\
     \midrule
     \textsc{trsm} & Solve $\Op(A) * X = \alpha * B$ or $X * \Op(A) = \alpha * B$, where $A$ is triangular and $B$ is general. $B$ is overwritten by the solution matrix X & $m^2n$ if $A$ is on the left; otherwise, $mn^2$ & $\Op(L^{-1})\Op(G)$, $\Op(G) \Op(L^{-1})$ \\
     \midrule
     \textsc{trsysv} & Solve $\Op(A) * X = B$ or $X * \Op(A) = B$, where $A$ is triangular and $B$ is symmetric. $B$ is overwritten by the solution matrix X & $m^3$ & $\Op(L^{-1})S$, $S\Op(L^{-1})$ \\
     \midrule
     \textsc{trtrsv} & Solve $\Op(A) * X = \alpha * B$ or $X * \Op(A) = \alpha * B$, where $A$ and $B$ are triangular. $B$ is overwritten by the solution matrix X & $m^3/3$ is $\Op(A)$ and $B$ have the same triangularity; otherwise $m^3$ & $\Op(L_1^{-1}) \Op(L_2)$, $\Op(L_2) \Op(L_1^{-1})$ \\
     \bottomrule
    \end{tabular}
\end{table*}